\documentclass[12pt,a4paper]{iopart}
\usepackage{iopams}
\usepackage{graphicx}
\usepackage[normalem]{ulem}

\eqnobysec

\newcommand{\ee}{\mathrm{e}}
\newcommand{\ii}{\mathrm{i}}

\def\nn{\nonumber}

\def\sgn{\mathrm{sgn}}

\def\eqref#1{(\ref{#1})}
\newtheorem{thm}{Theorem}[section]
\newtheorem{prop}[thm]{Proposition}

\begin{document}

\title[Exact time-evolving scattering states in open quantum-dot systems]
{Exact time-evolving scattering states\\
in open quantum-dot systems with an interaction:\\
Discovery of time-evolving resonant states}

\author{Akinori Nishino$^1$ and Naomichi Hatano$^2$}

\address{$^1$ Faculty of Engineering, Kanagawa University, 
3-27-1 Rokkakubashi, Kanagawa-ku, Yokohama, Kanagawa, 221-8686, Japan}
\address{$^2$ Institute of Industrial Science, The University of Tokyo, 
5-1-5 Kashiwanoha, Kashiwa, Chiba 277-8574, Japan}
\ead{nishino@kanagawa-u.ac.jp}
\vspace{10pt}
\begin{indented}
\item[]November 2023
\end{indented}

\begin{abstract}
We study exact time-evolving many-electron states
of an open double quantum-dot system with an interdot Coulomb interaction.
A systematic construction of the time-evolving states 
for arbitrary initial conditions is proposed.
For any initial states of 
one- and two-electron plane waves on the electrical leads, 
we obtain exact solutions of the time-evolving scattering states, 
which converge to known stationary scattering eigenstates in the long-time limit.
For any initial states of localized electrons on the quantum dots,
we find exact time-evolving states of a new type, 
which we refer to as time-evolving resonant states.
In contrast to stationary resonant states, whose wave functions
spatially diverge and not normalizable,
the time-evolving resonant states are normalizable
since their wave functions are restricted to a finite space interval 
due to causality.
The exact time-evolving resonant states enable us to
calculate the time-dependence of the survival probability of electrons on the quantum dots
for the system with the linearized dispersions.
It decays exponentially in time 
on one side of an exponential point of resonance energies
while, on the other side, it oscillates during the decay 
as a result of the interference of the two resonance energies.
\end{abstract}

%
\vspace{2pc}

\submitto{\jpa}
%
%
%


\section{Introduction}

Quantum transport in open quantum-dot systems with interacting electrons
has been intensively studied in recent years. 
One of the pioneering experiments is a measurement of zero-bias resonance peaks of 
electrical conductance across a quantum dot at low temperatures,
which is understood as the nonequilibrium Kondo effect~\cite{%
Cronenwett-Oosterkamp-Kouwenhoven_98Science,%
GoldhaberGordon-Shtrikman-Mahalu-Abusch-Magder-Meirav-Kastner_98Nature,%
VanDerWiel-DeFranceschi-Fujisawa-Elzerman-Tarucha-Kouwenhoven_00Science}.
On the other hand, in order to analyze the quantum transport 
in open quantum-dot systems theoretically, 
it is necessary to consider nonequilibrium states realized in the systems. 
The Landauer formula, which was first developed in noninteracting cases,
assumes that an electric current in coherent electron transport
is carried by quantum-mechanical scattering states~\cite{%
Landauer_57IBMJRD,Buttiker_86PRL,Bagwell-Orlando_98PRB}. 
The non-equilibrium Green's-function method has been successful
as an extension of the Landauer formula to interacting cases~\cite{%
Meir-Wingreen_92PRL,Meir-Wingreen-Lee_93PRL,Wingreen-Meir_94PRB,Costi-Hewson-Zlatic_94JPCM}. 

However, explicit many-electron wave functions realized in open quantum-dot systems
had not been obtained in interacting cases,
until we constructed {\it exact} many-electron scattering eigenstates 
for an open single quantum-dot system described by the interacting resonant-level model~\cite{%
Nishino-Imamura-Hatano_09PRL,Nishino-Imamura-Hatano_11PRB,Nishino-Hatano-Ordonez_15PRB}. 
We found it remarkable that incident free-electron plane waves are partially scattered to
{\it many-body bound states} around the quantum-dot due to the Coulomb interaction. 
By using the exact solution, we calculated the average electric current 
in the system under finite bias voltages, 
which agrees with the results by other methods~\cite{Doyon_07PRL,%
Boulat-Saleur-Schmitteckert_08PRL,Golub_07PRB,%
Karrasch-Andergassen-Pletyukhov-Schuricht-Borda-Meden-Schoeller_10EL}. 
Besides the interacting resonant-level model,
we also constructed exact scattering eigenstates
for the Anderson model~\cite{Imamura-Nishino-Hatano_09PRB} and 
the open double quantum-dot systems~\cite{Nishino-Imamura-Hatano_12JPC,Nishino-Hatano-Ordonez_16JPC}. 

In open quantum systems, resonant states with complex energy eigenvalues
have attracted renewed interest. 
The concept of the resonant state was originally introduced 
for the study of decaying states of unstable nuclei;
the imaginary part of the complex energy eigenvalue
corresponds to the lifetime of the decaying state~\cite{%
Gamow_28ZPhysA,Siegert_39PR,Peierls_59PRSLA}. 
However, the resonant state is often regarded as ``unphysical'' 
since its wave function diverges in space and is not normalizable. 
Normalization and probabilistic interpretation of the resonant state have been
discussed for many years~\cite{Hokkyo_65PTP,Berggren_68NPA,Romo_68NPA,%
Berggren_70PLB,Lind_93PRA,Hatano-Sasada-Nakamura-Petrosky_08PTP}.
Recently, resonant states are employed for understanding 
the resonant transport in open quantum-dot systems~\cite{%
Hatano-Sasada-Nakamura-Petrosky_08PTP,Hatano-Ordonez_19Book,Hatano-Ordonez_19Entropy}.
The two-body bound state
appearing in the exact many-electron scattering states~\cite{%
Nishino-Imamura-Hatano_09PRL,Nishino-Imamura-Hatano_11PRB,Nishino-Hatano-Ordonez_15PRB,%
Imamura-Nishino-Hatano_09PRB,Nishino-Imamura-Hatano_12JPC,Nishino-Hatano-Ordonez_16JPC} 
is regarded as a pair of two resonant states,
where the imaginary part of the complex energy eigenvalue 
characterizes the binding strength of the two-body bound states.

In the present article, we study time-evolving many-electron states 
of an open double quantum-dot system of spinless electrons 
with an interdot Coulomb interaction.
We analyze a general double quantum-dot system of 
arbitrary arrangements of two quantum dots including
serial, parallel, and T-shaped double quantum-dot systems~\cite{Tanaka-Kawakami_05PRB}.
Quite recently, for the open quantum systems with a single quantum dot,
exact time-evolving many-electron scattering states 
have been constructed for initial states of free-electron plane waves
under an assumption on the form of wave functions~\cite{%
Culver-Andrei_21PRB_1,Culver-Andrei_21PRB_2,Culver-Andrei_21PRB_3}.
We develop a systematic construction of the time-evolving states 
for the open double quantum-dot system
under {\it arbitrary} initial conditions without
assuming the form of wave functions as was used in Refs.~\cite{%
Culver-Andrei_21PRB_1,Culver-Andrei_21PRB_2,Culver-Andrei_21PRB_3}.
For demonstrations, we explicitly construct exact solutions of the time-evolving states
for two types of initial states: 
i) the one- and two-electron plane-wave states on the electrical leads;
ii) the localized states on the two quantum dots. 
In the case i), we obtain the time-evolving scattering states,
which exponentially converge to the known stationary scattering 
eigenstates~\cite{Nishino-Hatano-Ordonez_16JPC} in the long-time limit. 
In particular, at {\it exceptional points} of the resonance energies, 
we find exponential functions multiplied by a linear term of time in the wave functions.
Here we remark that the purely exponential behavior without any deviations 
in the form of power-law behavior~\cite{%
Khalfin_68PZETF,Chiu-Sudarshan-Misra_77PRD,Petrosky-Tasaki-Prigogine_91PhysicaA,%
Petrosky-Ordonez-Prigogine_01PRA,Garmon-Petrosky-Simine-Segal_13FortschrPhys,%
Chakraborty-Sensarma_18PRB,Garmon-Noba-Ordonez-Segal_19PRD}
is due to the unbounded linear dispersion relation that we assume for the electrical leads.

In the one-electron case of the case ii), we find a new time-evolving state,
which we refer to as a {\it time-evolving resonant state}.
The wave function of the time-evolving resonant state increases exponentially
only inside a finite space interval and hence is normalizable, 
in contrast to the ``unphysical'' resonant states
with spatially diverging wave functions~\cite{Hatano-Sasada-Nakamura-Petrosky_08PTP}.
The restriction of the wave function to the space interval is originated from causality,
which was also observed in the Friedrichs model~\cite{Petrosky-Ordonez-Prigogine_01PRA}.
We show through the construction of exact solutions
that the wave function coincides with the resonant state
inside the space interval.
Furthermore, in the two-electron case of the case ii), 
we obtain a {\it time-evolving two-body resonant state}
that corresponds to the two-body resonance pole in the interacting case.

The explicit wave functions of time-evolving resonant states enable us to calculate
exactly the time-dependence of
the existence probability of electrons on the leads
and the survival probability of electrons on the quantum dots
under the initial states of localized electrons on the quantum dots.
The exponential behavior of the wave functions 
leads to that of both the existence and the survival probabilities.
It is remarkable that their behavior changes 
around exceptional points of the resonance energies.
In a parameter region on one side of an exceptional point,
we observe a pure exponential increase of the existence probability in space
and a pure exponential decay of the survival probability in time.
In a region on the other side of the exceptional point,
we observe oscillation during the exponential behavior,
which are results of the interference of two one-body resonance poles.
 
We also remark that the open double quantum-dot system is considered to be a charge-qubit device
that consists of one qubit and two quantum probes.
The two initial states above are related to the following two ways of manipulating
the device: the case i) can be utilized as ``initialization'' of the qubit
and the case ii) may correspond to ``coherent manipulation''~\cite{Chowdhury-Chattopadhyay_23AQT}.
The decoherence of qubit states after these ways of manipulation
is described by the time evolution of the quantum states under the initial conditions.
Our exact solutions of the time-evolving many-electron states
will be useful for calculating the decoherence time.

The paper is organized as follows: 
In Section~\ref{sec:openDQD}, 
an open double quantum-dot system with an interdot Coulomb interaction is introduced.
In Section~\ref{sec:time-evolving-scattering-states},
we propose a systematic construction of time-evolving states
under arbitrary initial conditions.
We construct exact solutions of the time-evolving scattering states
for any initial states of one- and two-electron plane waves on the electrical leads.
In Section~\ref{sec:time-evolving-resonant-states},
we obtain exact time-evolving resonant states for the initial states of 
one and two electrons localized on the quantum dots.
By using the exact solutions, we investigate
the existence probability of electrons on the electrical leads
and the survival probability of electrons on the quantum dots.
Section~\ref{sec:concluding-remarks} is devoted to concluding remarks.
Appendix provides a concise review of resonant states.

\section{Open double quantum-dot systems}
\label{sec:openDQD}

\begin{figure}[t]
\begin{center}
{
\begin{picture}(250,135)(0,0)
 \put(0,0){\includegraphics[width=220pt]{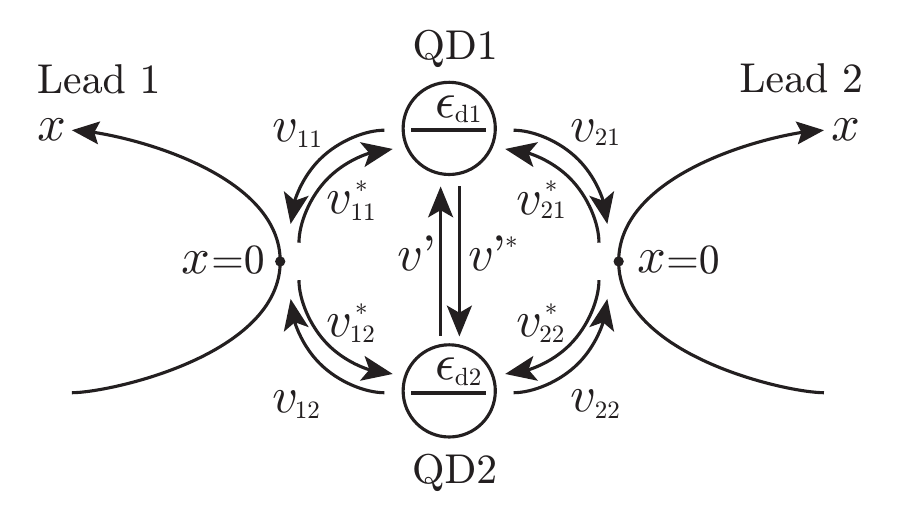}}
\end{picture}
}
\end{center}
\caption{
A schematic diagram of the double quantum-dot system.
}
\label{fig:DQD}
\end{figure}

We introduce an open double quantum-dot system with 
an interdot Coulomb interaction~\cite{Nishino-Hatano-Ordonez_16JPC}.
Let us consider an open quantum system which consists of
two quantum dots and two one-dimensional electrical leads of noninteracting spinless electrons
as is illustrated in Fig.~\ref{fig:DQD}.
We assume a single energy level for each quantum dot
and linearize the dispersion relation in the vicinity of the Fermi energy for each lead.
The two quantum dots are connected to each other
and  to the origin $x=0$ of each lead.
The Hamiltonian of the system is given by
\begin{eqnarray}
\label{eq:Hamiltonian_GDQD}
H=&\sum_{m=1,2}
   \int dx\,
   c^{\dagger}_{m}(x)
   v_{\rm F}\frac{1}{\ii}\frac{d}{dx}c_{m}(x)
  +\sum_{m=1,2 \atop \alpha=1,2}
   \big(v_{m\alpha}c^{\dagger}_{m}(0)d_{\alpha}
  +v_{m\alpha}^{\ast}d^{\dagger}_{\alpha}c_{m}(0)\big) 
  \nn\\
 &+v^{\prime}d^{\dagger}_{1}d_{2}+v^{\prime\ast}d^{\dagger}_{2}d_{1}
  +\sum_{\alpha=1,2}\epsilon_{{\rm d}\alpha}n_{{\rm d}\alpha}
  +U^{\prime}n_{{\rm d}1}n_{{\rm d}2}. 
\end{eqnarray}
Here $c^{\dagger}_{m}(x)$ and $c_{m}(x)$ are
the creation and annihilation operators of electrons 
at the position $x$ on the lead $m$, respectively,
and $d^{\dagger}_{\alpha}$ and $d_{\alpha}$ are those on the quantum dot $\alpha$.
We also define the number operator $n_{{\rm d}\alpha}=d^{\dagger}_{\alpha}d_{\alpha}$
of electrons on the quantum dot $\alpha$.
The parameter $v_{m\alpha}$ is 
the transfer integral between the lead $m$ and the quantum dot $\alpha$,
while $v^{\prime}$ is that between the two quantum dots.
Here we assume the S-wave scattering of electrons off the quantum dots,
and hence the transfer integrals $v_{m\alpha}$ and $v^{\prime}$ are independent 
of the wave numbers of electrons.
The parameter $\epsilon_{{\rm d}\alpha}$ is the energy level on the quantum dot $\alpha$
and $U^{\prime}$ expresses the strength of the interdot Coulomb interaction.
Here and hereafter, we set $\hbar=1$.
In what follows, the Fermi velocity $v_{\rm F}$ of the leads is also set to unity: $v_{\rm F}=1$.

We remark that actual energy levels on the quantum dots
are affected by the electrons contained in the quantum dots.
In order to investigate the electron distribution 
within a quantum dot of finite size, we would have to solve
the Schr\"odinger equations and the Poisson equation 
self-consistently~\cite{Chowdhury-Chattopadhyay_23AQT},
which, however, is out of scope of the present paper.

The Hamiltonian in Eq.~(\ref{eq:Hamiltonian_GDQD})
includes various arrangements of the two quantum dots~\cite{Tanaka-Kawakami_05PRB}.
For example, in the case of $v^{\prime}=0$, we have a parallel-coupled double quantum-dot system;
in the case of $v_{12}=v_{21}=0$, we have a serial-coupled double quantum-dot system;
in the case of $v_{12}=v_{22}=0$, we have a T-shaped double quantum-dot system.
The arrangement of the two quantum dots affects
the quantum interference of electrons traveling through the quantum dots.

Let us derive the time-dependent Schr\"odinger equation 
for the open double quantum-dot system in Eq.~\eqref{eq:Hamiltonian_GDQD}.
First, we consider the one-electron case, 
which is equivalent to the noninteracting case.
The time-dependent one-electron state is given in the form
\begin{eqnarray}
 |\Psi(t)\rangle
 =\Big(\int\!dx
  \sum_{m=1, 2}g_{m}(x,t)c^{\dagger}_{m}(x)
  +\sum_{\alpha=1, 2}e_{\alpha}(t)d^{\dagger}_{\alpha}\Big)|0\rangle.
\end{eqnarray}
Here $|0\rangle$ is the vacuum state.
The wave functions $g_{m}(x,t)$ for $m=1, 2$ and $e_{\alpha}(t)$ for $\alpha=1,2$
are determined by the time-dependent Schr\"odinger equation
$\ii\partial_{t}|\Psi(t)\rangle=H|\Psi(t)\rangle$,
which is explicitly written in the form of the following coupled partial-differential equations:
\numparts
\label{eq:Sch-eq_1-state_GDQD}
\begin{eqnarray}
\label{eq:Sch-eq_1-state_GDQD_g}
&\ii\partial_{t}g_{m}(x,t)
 =\frac{1}{\ii}\partial_{x}g_{m}(x,t)
 +\delta(x)\sum_{\alpha}v_{m\alpha}e_{\alpha}(t), \\
\label{eq:Sch-eq_1-state_GDQD_e}
&\ii\partial_{t}e_{\alpha}(t)
 =\epsilon_{{\rm d}\alpha}e_{\alpha}(t)
 +\sum_{m}v_{m\alpha}^{\ast}g_{m}(0,t)
 +v^{\prime}_{\alpha}e_{\overline{\alpha}}(t).
\end{eqnarray}
\endnumparts
Here $\delta(x)$ is Dirac's delta function.
We expressed the partial differential operators $\partial/\partial t$ and $\partial/\partial x$ 
by $\partial_{t}$ and $\partial_{x}$, respectively, 
and set $\overline{\alpha}=3-\alpha$,
$v_{1}^{\prime}=v^{\prime}$ and $v_{2}^{\prime}=v^{\prime\ast}$.
We remark that the time-dependent states in noninteracting cases were 
studied for the single-quantum-dot systems 
with general dispersion relations in Ref.~\cite{Gurvitz_15PhysScr}.

Next, we consider the two-electron case. 
The time-dependent two-electron state is given in the form
\begin{eqnarray}
\label{eq:2-state_GDQD}
 |\Psi(t)\rangle
&=\Big(\int_{x_{1}<x_{2}}\hspace{-15pt}dx_{1}dx_{2}
  \sum_{m_{1}, m_{2}=1, 2}g_{m_{1}m_{2}}(x_{1},x_{2},t)
  c^{\dagger}_{m_{1}}(x_{1})c^{\dagger}_{m_{2}}(x_{2})
  \nn\\
&\quad
  +\sum_{m=1, 2}\sum_{\alpha=1, 2}e_{m,\alpha}(t)c^{\dagger}_{m}(x)d^{\dagger}_{\alpha}
  +f_{12}(t)d^{\dagger}_{1}d^{\dagger}_{2}\Big)|0\rangle,
\end{eqnarray}
where we assume the following Fermionic anti-symmetries
for the wave functions $g_{m_{1}m_{2}}(x_{1},x_{2},t)$ and $f_{12}(t)$:
\begin{eqnarray}
 \label{eq:2-state_GDQD_anti-sym}
 g_{m_{1}m_{2}}(x_{1},x_{2},t)=-g_{m_{2}m_{1}}(x_{2},x_{1},t),\qquad
 f_{12}(t)=-f_{21}(t).
\end{eqnarray}
The time-dependent Schr\"odinger equation 
$\ii\partial_{t}|\Psi(t)\rangle=H|\Psi(t)\rangle$ gives
the following coupled partial-differential equations for the wave functions 
$g_{m_{1}m_{2}}(x_{1},x_{2},t)$, $e_{m,\alpha}(x,t)$ and $f_{\alpha\overline{\alpha}}(t)$:
\numparts
\begin{eqnarray}
\label{eq:Sch-eq_2-state_GDQD_g}
 \ii\partial_{t}g_{m_{1}m_{2}}(x_{1},x_{2},t)
 =\frac{1}{\ii}(\partial_{x_{1}}\!+\!\partial_{x_{2}})
  g_{m_{1}m_{2}}(x_{1},x_{2},t)
 \nn\\
 +\sum_{\alpha}\big(v_{m_{2}\alpha}\delta(x_{2})e_{m_{1},\alpha}(x_{1},t)
  \!-\! v_{m_{1}\alpha}\delta(x_{1})e_{m_{2},\alpha}(x_{2},t)\big), 
  \\
\label{eq:Sch-eq_2-state_GDQD_e}
 \ii\partial_{t}e_{m,\alpha}(x,t)
 =\Big(\frac{1}{\ii}\partial_{x}+\epsilon_{{\rm d}\alpha}\Big)e_{m,\alpha}(x,t)
 \nn\\
 +\sum_{n}v_{n\alpha}^{\ast}g_{mn}(x,0,t)
 +v^{\prime}_{\alpha}e_{m,\overline{\alpha}}(x,t)
 -v_{m\overline{\alpha}}\delta(x)f_{\alpha\overline{\alpha}}(t),
 \\
\label{eq:Sch-eq_2-state_GDQD_f}
 \ii\partial_{t}f_{\alpha\overline{\alpha}}(t)
 =(2\overline{\epsilon}_{\rm d}+U^{\prime})f_{\alpha\overline{\alpha}}(t)
  +\sum_{m}\big(v^{\ast}_{m\alpha}e_{m,\overline{\alpha}}(0,t)
  -v^{\ast}_{m\overline{\alpha}}e_{m,\alpha}(0,t)\big).
\end{eqnarray}
\endnumparts
Here we introduced $\overline{\epsilon}_{\rm d}=(\epsilon_{{\rm d}1}+\epsilon_{{\rm d}2})/2$.

The many-electron scattering eigenstates have been constructed
as the stationary solutions of the set of Schr\"odinger equations in 
Eqs.~\eqref{eq:Sch-eq_2-state_GDQD_g}, \eqref{eq:Sch-eq_2-state_GDQD_e}
and \eqref{eq:Sch-eq_2-state_GDQD_f} under the scattering boundary 
conditions~\cite{Nishino-Imamura-Hatano_12JPC,Nishino-Hatano-Ordonez_16JPC}.
As an effect of the interdot interaction, 
the incident plane waves are partially scattered to 
two- and three-body bound states at the quantum dots.
The binding strength of the many-body bound states corresponds to
the imaginary part of the resonance poles of the scattering wave functions.

\mathversion{bold}
\section{Time-evolving scattering states}
\mathversion{normal}
\label{sec:time-evolving-scattering-states}

\mathversion{bold}
\subsection{One-electron scattering states}
\mathversion{normal}
\label{sec:1-time-evolving-scattering-states}

Let us construct time-evolving scattering states of the open double quantum-dot system
by solving initial-value problems for the time-dependent Schr\"odinger equation 
given in Sec.~\ref{sec:openDQD}.
We propose a systematic construction of exact solutions
for arbitrary initial states at time $t=0$
without assuming the form of wave functions as was used in Refs.~\cite{%
Culver-Andrei_21PRB_1,Culver-Andrei_21PRB_2,Culver-Andrei_21PRB_3}.
In this subsection, we consider the one-electron case.
The Schr\"odinger equation~\eqref{eq:Sch-eq_1-state_GDQD_g} tells us that,
in each region of $x>0$ or $x<0$, the general solution $g_{m}(x,t)$ 
is given by an arbitrary function $F(x-t)$ of the variable $x-t$.
In other words, for the real parameter $a$ under the condition
$x+a>0$ if $x>0$ or the condition $x+a<0$ if $x<0$,
the solution $g_{m}(x,t)$ is invariant under the translation $(x, t)\mapsto (x+a, t+a)$.
By combining the two conditions as $x(x+a)>0$, we have
\begin{eqnarray}
\label{eq:Sch-eq_1-state_GDQD_g_2}
 g_{m}(x+a,t+a)=g_{m}(x,t),\quad \mbox{if } x(x+a)>0.
\end{eqnarray}

Because of the delta-function term in Eq.~\eqref{eq:Sch-eq_1-state_GDQD_g},
the wave function $g_{m}(x,t)$ is discontinuous at $x=0$.
Hence, at $x=0$, we match in the following form
the wave functions that are given in the two regions $x>0$ and $x<0$.
We obtain the matching condition of $g_{m}(x,t)$ at $x=0$
by integrating Eq.~\eqref{eq:Sch-eq_1-state_GDQD_g}
over an infinitesimal region around $x=0$, which results in
\begin{eqnarray}
\label{eq:Sch-eq_1-state_GDQD_g_3}
 g_{m}(0+,t)-g_{m}(0-,t)=-\ii\sum_{\alpha}v_{m\alpha}e_{\alpha}(t).
\end{eqnarray}
Since the value of $g_{m}(x,t)$ at $x=0$, 
which appears in Eq.~\eqref{eq:Sch-eq_1-state_GDQD_e}, is not determined 
by the Schr\"odinger equations~\eqref{eq:Sch-eq_1-state_GDQD_g} and \eqref{eq:Sch-eq_1-state_GDQD_e},
we assume 
\begin{eqnarray}
\label{eq:Sch-eq_1-state_GDQD_g_4}
 g_{m}(0,t)=\frac{1}{2}\big(g_{m}(0+,t)+g_{m}(0-,t)\big).
\end{eqnarray}
We remark that the difference of the value $g_{m}(0,t)$ 
should be absorbed into the system parameters
by the renormalization-group method~\cite{Hewson}.

By inserting Eqs.~\eqref{eq:Sch-eq_1-state_GDQD_g_4} and \eqref{eq:Sch-eq_1-state_GDQD_g_3}
into Eq.~\eqref{eq:Sch-eq_1-state_GDQD_e}, we obtain
\begin{eqnarray}
\fl
 \ii\partial_{t}e_{\alpha}(t)
&=(\epsilon_{{\rm d}\alpha}-\ii\Gamma_{\alpha\alpha})e_{\alpha}(t)
  +\big(v^{\prime}_{\alpha}-\ii\Gamma_{\alpha\overline{\alpha}}\big)e_{\overline{\alpha}}(t)
  +\sum_{m}v_{m\alpha}^{\ast}g_{m}(0-,t).
\label{eq:Sch-eq_1-state_GDQD_e_1}
\end{eqnarray}
Here we introduced the level width
$\Gamma_{\alpha\beta}=\sum_{m}v^{\ast}_{m\alpha}v_{m\beta}/2$.
By introducing an auxiliary function $\tilde{e}_{\alpha}(t)
=\ee^{\ii(\epsilon_{{\rm d}\alpha}-\ii\Gamma_{\alpha\alpha})t}e_{\alpha}(t)$, 
the differential equation~\eqref{eq:Sch-eq_1-state_GDQD_e_1} is rewritten as
\begin{eqnarray}
\fl
 \ii\partial_{t}\tilde{e}_{\alpha}(t)
 &=(v^{\prime}_{\alpha}-\ii\Gamma_{\alpha\overline{\alpha}})
  \ee^{\ii(\Delta\epsilon_{{\rm d}\alpha}-\ii\Delta\Gamma_{\alpha})t}\tilde{e}_{\overline{\alpha}}(t)
   +\sum_{m}v_{m\alpha}^{\ast}
  \ee^{\ii(\epsilon_{{\rm d}\alpha}-\ii\Gamma_{\alpha\alpha})t}g_{m}(0-,t),
\end{eqnarray}
where $\Delta\epsilon_{{\rm d}\alpha}=\epsilon_{{\rm d}\alpha}-\epsilon_{{\rm d}\overline{\alpha}}$ 
and $\Delta\Gamma_{\alpha}=\Gamma_{\alpha\alpha}-\Gamma_{\overline{\alpha}\,\overline{\alpha}}$.

The coupled first-order differential equations for $\tilde{e}_{\alpha}(t)$ 
and $\tilde{e}_{\overline{\alpha}}(t)$
are transformed to a second-order differential equation for $\tilde{e}_{\alpha}(t)$ 
of the form
\begin{eqnarray}
\fl
&\partial^{2}_{t}\tilde{e}_{\alpha}(t)
 -\ii(\Delta\epsilon_{{\rm d}\alpha}-\ii\Delta\Gamma_{\alpha})
  \partial_{t}\tilde{e}_{\alpha}(t)
 +(v^{\prime}_{\alpha}-\ii\Gamma_{\alpha\overline{\alpha}})
  (v^{\prime}_{\overline{\alpha}}-\ii\Gamma_{\overline{\alpha}\alpha})
  \tilde{e}_{\alpha}(t)
  =G_{\alpha}(t),
 \label{eq:Sch-eq_1-state_GDQD_e_2}
\end{eqnarray}
where the inhomogeneous term is given by
\begin{eqnarray}
 \label{eq:Sch-eq_1-state_GDQD_G}
\fl
 G_{\alpha}(t)
  =-\sum_{m}\ee^{\ii(\epsilon_{{\rm d}\alpha}-\ii\Gamma_{\alpha\alpha})t}
 \bigg[v_{m\alpha}^{\ast}
  (\ii\partial_{t}-\epsilon_{{\rm d}\overline{\alpha}}+\ii\Gamma_{\overline{\alpha}\,\overline{\alpha}})
  +v_{m\overline{\alpha}}^{\ast}
  (v^{\prime}_{\alpha}-\ii\Gamma_{\alpha\overline{\alpha}})\bigg]
 g_{m}(0-,t).
\end{eqnarray}
In order to solve Eq.~\eqref{eq:Sch-eq_1-state_GDQD_e_2},
we investigate the characteristic equation of a homogeneous linear differential equation
associated with Eq.~\eqref{eq:Sch-eq_1-state_GDQD_e_2} as 
\begin{eqnarray}
&\lambda^{2}
  -\ii(\Delta\epsilon_{{\rm d}\alpha}-\ii\Delta\Gamma_{\alpha})\lambda
  +(v^{\prime}_{\alpha}-\ii\Gamma_{\alpha\overline{\alpha}})
     (v^{\prime}_{\overline{\alpha}}-\ii\Gamma_{\overline{\alpha}\alpha})=0,
\end{eqnarray}
which gives the two characteristic roots as
\begin{eqnarray}
\label{eq:lambda-eta}
&\lambda_{\alpha,\pm}
  =\frac{1}{2}\big(\ii(\Delta\epsilon_{{\rm d}\alpha}-\ii\Delta\Gamma_{\alpha})+\eta_{\pm}\big),
  \nn\\
&\eta_{\pm}=\pm\eta=\pm\sqrt{(\ii\Delta\epsilon_{{\rm d}\alpha}+\Delta\Gamma_{\alpha})^{2}
  -4(v^{\prime}-\ii\Gamma_{12})(v^{\prime\ast}-\ii\Gamma_{21})}.
\end{eqnarray}
We notice that the two roots $\lambda_{\alpha,\pm}$ merge into one 
at the parameter points giving $\eta=0$.

The general solution of Eq.~\eqref{eq:Sch-eq_1-state_GDQD_e_2} is obtained as
\begin{eqnarray}
\fl
\tilde{e}_{\alpha}(t)
&=\cases{
 \sum_{s=\pm}C_{\alpha,s}\ee^{\lambda_{\alpha,s}t}
  +\sum_{s=\pm}\frac{1}{\lambda_{\alpha,s}-\lambda_{\alpha,\overline{s}}}
    \int^{t}_{-\infty}\hspace{-8pt}dt^{\prime}\,\ee^{\lambda_{\alpha,s}(t-t^{\prime})}G_{\alpha}(t^{\prime})
 & for $\eta\neq 0$, \\
 (D_{\alpha,0}+tD_{\alpha, 1})\ee^{\lambda^{0}_{\alpha}t}
  +\int_{-\infty}^{t}\hspace{-5pt}dt^{\prime}\,(t-t^{\prime})
   \ee^{\lambda^{0}_{\alpha}(t-t^{\prime})}G_{\alpha}(t^{\prime})
 & for $\eta=0$, \\
 }
 \label{eq:Sch-eq_1-state_GDQD_e_3}
\end{eqnarray}
where $C_{\alpha,s}$ for $s=\pm$ 
and $D_{\alpha,r}$ for $r=0, 1$ are integration constants,
and $\lambda^{0}_{\alpha}=\lambda_{\alpha,s}|_{\eta=0}$.
For $\eta=0$ in Eq.~\eqref{eq:Sch-eq_1-state_GDQD_e_3}, 
we find terms linear in time $t$ inside and outside the integral with respect to the variable $t^{\prime}$.
We shall elucidate in \ref{sec:1-ResonantStates}
that the parameter points satisfying $\eta=0$
correspond to {\it exceptional points} of a non-Hermite effective Hamiltonian.

\begin{figure}[t]
\begin{center}
{
\begin{picture}(290,100)(0,0)
 \put(0,0){\includegraphics[width=270pt]{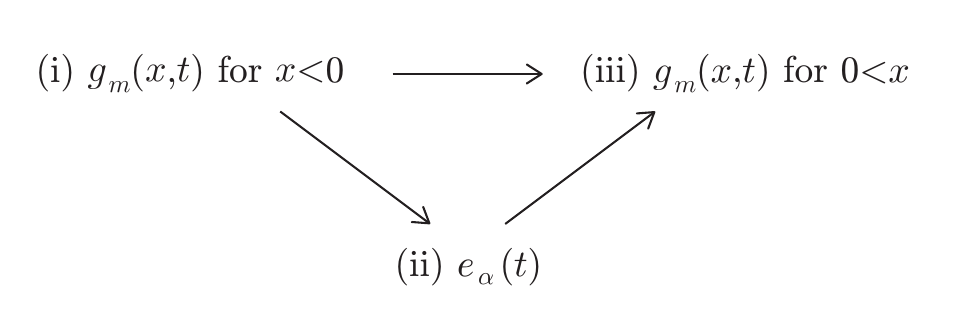}}
\end{picture}
}
\end{center}
\caption{The flow chart of the construction of the time-evolving one-electron states.}
\label{fig:flow-chart_1-elec}
\end{figure}

Now we propose a systematic construction of the time-evolving one-electron states 
under arbitrary initial conditions
as is shown by the flow chart in Fig.~\ref{fig:flow-chart_1-elec}:
\begin{description}
\itemsep=0pt
\item[Step 1)]
The wave function $g_{m}(x,t)$ for $x<0$ in (i) of Fig.~\ref{fig:flow-chart_1-elec}
is obtained by the translation invariance in Eq.~\eqref{eq:Sch-eq_1-state_GDQD_g_2}
with the wave function $g_{m}(x,0)$ of given initial states.
\item[Step 2)] 
In the process (i)$\to$(ii) in Fig.~\ref{fig:flow-chart_1-elec}, 
we use the general solution of $e_{\alpha}(t)$ in Eq.~\eqref{eq:Sch-eq_1-state_GDQD_e_3}.
The integration constant $C_{\alpha,s}$ or $D_{\alpha,r}$
is determined by the initial conditions for $e_{\alpha}(t)$.
\item[Step 3)]
In the process (i), (ii)$\to$(iii) in Fig.~\ref{fig:flow-chart_1-elec}, 
we use the matching condition of $g_{m}(x,t)$ in Eq.~\eqref{eq:Sch-eq_1-state_GDQD_g_3}.
\end{description}
We find that the construction above provides an exact solution 
if we obtain the integration in Eq.~\eqref{eq:Sch-eq_1-state_GDQD_e_3}, 
whose integrand is given by the initial state.

Let us now specifically consider the time-evolving scattering states for the initial conditions
\begin{eqnarray}
&g_{m}(x,0)=\frac{1}{\sqrt{2\pi}}\delta_{m,\ell}\ee^{\ii kx}\; \mbox{ for } m=1, 2, \quad
  e_{\alpha}(0)=0\; \mbox{ for } \alpha=1, 2,
\label{eq:1-initial-state_GDQD}
\end{eqnarray}
which stand for the one-electron plane-wave state
with the wave number $k$ on the entire lead $\ell$
along with the absence of electrons on the quantum dots.

\begin{prop}
The set of Schr\"odinger equations~\eqref{eq:Sch-eq_1-state_GDQD_g}
and \eqref{eq:Sch-eq_1-state_GDQD_e}
for the initial conditions~\eqref{eq:1-initial-state_GDQD} is solved as follows:
i) For $\eta\neq 0$, the solution is given by
\numparts
\begin{eqnarray}
\label{eq:1-scattering-state_GDQD_g}
\fl 
 g^{(\ell)}_{m,k}(x,t)
 =\frac{1}{\sqrt{2\pi}}\delta_{m,\ell}\ee^{\ii k(x-t)}
 \nn\\
\fl
 \hspace{60pt}
  \!+\!\sum_{\alpha, s}e^{(\ell)}_{\alpha,k}\, y_{m\overline{\alpha},\overline{s}}
  \Big[\ee^{\ii k(x-t)}
  \!-\!\ee^{\ii (\overline{\epsilon}_{{\rm d}}-\ii \overline{\Gamma}+\ii \eta_{s}/2)(x-t)}\Big]
  \tilde{\theta}(t\!-\! x)\theta(x),
 \\
\label{eq:1-scattering-state_GDQD_e}
\fl
 e^{(\ell)}_{\alpha,k}(t)
 =e^{(\ell)}_{\alpha,k}\ee^{-\ii k t}
  +\sum_{s=\pm}\frac{1}{\eta_{s}}
  \bigg[\lambda_{\alpha,\overline{s}}e^{(\ell)}_{\alpha,k}
  +\ii (v^{\prime}_{\alpha}-\ii \Gamma_{\alpha\overline{\alpha}})
  e^{(\ell)}_{\overline{\alpha},k}\bigg]
  \ee^{-\ii (\overline{\epsilon}_{{\rm d}}-\ii \overline{\Gamma}+\ii \eta_{s}/2)t},
\end{eqnarray}
\endnumparts
where we used
\begin{eqnarray}
\label{eq:1-eley_GDQD}
&e^{(\ell)}_{\alpha,k}=\frac{1}{\sqrt{2\pi}}
  \frac{(k-\epsilon_{{\rm d}\overline{\alpha}}+\ii \Gamma_{\overline{\alpha}\,\overline{\alpha}})v_{\ell\alpha}^{\ast}
  +(v^{\prime}_{\alpha}-\ii \Gamma_{\alpha\overline{\alpha}})v_{\ell\overline{\alpha}}^{\ast}}
  {(k-\overline{\epsilon}_{{\rm d}}+\ii \overline{\Gamma}-\ii \eta/2)
  (k-\overline{\epsilon}_{{\rm d}}+\ii \overline{\Gamma}+\ii \eta/2)},
 \nn\\
&y_{m\alpha,\pm}
 =\frac{1}{\eta_{s}}
  (\ii \lambda_{\alpha,\mp}v_{m\overline{\alpha}}
  \!+\!(v_{\alpha}^{\prime}\!\!-\!\ii \Gamma_{\alpha\overline{\alpha}})
  v_{m\alpha})
\end{eqnarray}
and the following two types of step functions:
\begin{equation}
\label{eq:step-func}
 \theta(t)=
 \cases{
  1 & if $t>0$, \\
  1/2 & if $t=0$, \\
  0 & if $t<0$, \\
 }
 \quad
 \tilde{\theta}(t)=
 \cases{
  1 & if $t>0$, \\
  0 & if $t\leq 0$. \\
 }
\end{equation}
ii) For $\eta=0$, the solution is given by
\numparts
\begin{eqnarray}
\label{eq:1-scattering-state_GDQD_g_ep}
\fl 
 g^{(\ell)}_{m,k}(x,t)
 =\frac{1}{\sqrt{2\pi}}\delta_{m,\ell}\ee^{\ii k(x-t)}
 -\ii \sum_{\alpha}v_{m\alpha}
  \Big[e^{(\ell)0}_{\alpha,k}\ee^{\ii k(x-t)}
 \nn\\
\fl
 \hspace{58pt}
  +\big((\lambda^{0}_{\alpha}e^{(\ell)0}_{\alpha,k}
  +\ii (v^{\prime}_{\alpha}-\ii \Gamma_{\alpha\overline{\alpha}})e^{(\ell)0}_{\overline{\alpha},k})(t\!-\! x)
  -e^{(\ell)0}_{\alpha,k}\big)
  \ee^{\ii (\overline{\epsilon}_{{\rm d}}-\ii \overline{\Gamma})(x-t)}
  \Big]\tilde{\theta}(t\!-\! x)\theta(x),
 \\
\label{eq:1-scattering-state_GDQD_e_ep}
\fl
 e^{(\ell)}_{\alpha,k}(t)
 =e^{(\ell)0}_{\alpha,k}\ee^{-\ii k t}
  +\Big[\big(\lambda^{0}_{\alpha}e^{(\ell)0}_{\alpha,k}
  +\ii (v^{\prime}_{\alpha}-\ii \Gamma_{\alpha\overline{\alpha}})e^{(\ell)0}_{\overline{\alpha},k}\big)t
  -e^{(\ell)0}_{\alpha,k}\Big]
  \ee^{-\ii (\overline{\epsilon}_{{\rm d}}-\ii \overline{\Gamma})t},
\end{eqnarray}
\endnumparts
where $\lambda^{0}_{\alpha}=\lambda_{\alpha,s}|_{\eta=0}$
and $e^{(\ell)0}_{\alpha,k}=e^{(\ell)}_{\alpha,k}|_{\eta=0}$.
\end{prop}
\noindent
{\it Proof.}
By following the three steps above, we show a proof in the case $\eta\neq 0$.

\noindent
{\bf Step 1)}  We obtain the wave function $g_{m}(x,t)$ in the region of $x<0$ for $t>0$ 
through the translation invariance in Eq.~\eqref{eq:Sch-eq_1-state_GDQD_g_2} as
\begin{eqnarray}
 g_{m}(x,t)=g_{m}(x-t,0)=\frac{1}{\sqrt{2\pi}}\delta_{m,\ell}\ee^{\ii k(x-t)}
\end{eqnarray}
due to the relation $x(x-t)>0$.

\noindent
{\bf Step 2)} Process (i)$\to$(ii): By setting $x=0-$ in the above equation, we have
\begin{eqnarray}
g_{m}(0-,t)=\frac{1}{\sqrt{2\pi}}\delta_{m,\ell}\ee^{-\ii kt}.
\end{eqnarray}
Then the inhomogeneous term $G_{\alpha}(t)$ 
in Eq.~\eqref{eq:Sch-eq_1-state_GDQD_G} is calculated as
\begin{eqnarray}
 G_{\alpha}(t)
 =\frac{-1}{\sqrt{2\pi}}
  \ee^{\ii (-k+\epsilon_{{\rm d}\alpha}-\ii \Gamma_{\alpha\alpha})t}
  \bigg[v_{\ell\alpha}^{\ast}
  \big(k-\epsilon_{{\rm d}\overline{\alpha}}+\ii \Gamma_{\overline{\alpha}\,\overline{\alpha}})
  \!+\!v_{\ell\overline{\alpha}}^{\ast}
  (v^{\prime}_{\alpha}\!-\!\ii \Gamma_{\alpha\overline{\alpha}})\bigg].
\end{eqnarray}
By inserting this into the general solution in Eq.~\eqref{eq:Sch-eq_1-state_GDQD_e_3}, 
we obtain
\begin{eqnarray}
&\tilde{e}_{\alpha}(t)
 =\sum_{s}C_{\alpha,s}\ee^{\lambda_{\alpha,s}t}
  +e^{(\ell)}_{\alpha,k}\ee^{\ii (-k+\epsilon_{{\rm d}\alpha}-\ii \Gamma_{\alpha\alpha})t},
  \nn\\
\therefore\quad
&e_{\alpha}(t)
 =\ee^{-\ii (\epsilon_{{\rm d}\alpha}-\ii \Gamma_{\alpha\alpha})t}\tilde{e}_{\alpha}(t)
 =\sum_{s}C_{\alpha,s}\ee^{-\ii (\overline{\epsilon}_{{\rm d}}-\ii \overline{\Gamma}+\ii \eta_{s}/2)t}
  +e^{(\ell)}_{\alpha,k}\ee^{-\ii kt}.
 \label{eq:Sch-eq_1-state_GDQD_e_4}
\end{eqnarray}
Here we used the relation
$\epsilon_{{\rm d}\alpha}-\ii \Gamma_{\alpha\alpha}+\ii \lambda_{\alpha,s}
=\overline{\epsilon}_{{\rm d}}-\ii \overline{\Gamma}+\ii \eta_{s}/2$.
By applying the solution~\eqref{eq:Sch-eq_1-state_GDQD_e_4} to Eq.~\eqref{eq:Sch-eq_1-state_GDQD_e_1},
we obtain
\begin{eqnarray}
&\tilde{e}_{\overline{\alpha}}(t)
=\sum_{s}\frac{\ii \lambda_{\alpha,s}}{v^{\prime}_{\alpha}-\ii \Gamma_{\alpha\overline{\alpha}}}
  C_{\alpha,s}
  \ee^{\ii (\Delta\epsilon_{{\rm d}\overline{\alpha}}-\ii \Delta\Gamma_{\overline{\alpha}}-\ii \lambda_{\alpha,s})t}
  +e^{(\ell)}_{\overline{\alpha},k}
  \ee^{\ii (-k+\epsilon_{{\rm d}\overline{\alpha}}-\ii \Gamma_{\overline{\alpha}\,\overline{\alpha}})t},
  \nn\\
\therefore\quad
&e_{\overline{\alpha}}(t)
 =\sum_{s}\frac{\ii \lambda_{\alpha,s}}{v^{\prime}_{\alpha}-\ii \Gamma_{\alpha\overline{\alpha}}}
  C_{\alpha,s}\ee^{-\ii (\overline{\epsilon}_{{\rm d}}-\ii \overline{\Gamma}+\ii \eta_{s}/2)t}
  +e^{(\ell)}_{\overline{\alpha},k}\ee^{-\ii kt}.
 \label{eq:Sch-eq_1-state_GDQD_e_5}
\end{eqnarray}
The initial conditions $e_{\alpha}(0)=e_{\overline{\alpha}}(0)=0$
determine the integration constants $C_{\alpha,+}$ and $C_{\alpha,-}$ as
\begin{eqnarray}
&e_{\alpha}(0+)
  =C_{\alpha,+}+C_{\alpha,-}+e^{(\ell)}_{\alpha,k}=0,
  \nn\\
&e_{\overline{\alpha}}(0+)
=\frac{\ii \lambda_{\alpha,+}}{v^{\prime}_{\alpha}-\ii \Gamma_{\alpha\overline{\alpha}}}C_{\alpha,+}
  +\frac{\ii \lambda_{\alpha,-}}{v^{\prime}_{\alpha}-\ii \Gamma_{\alpha\overline{\alpha}}}C_{\alpha,-}
  +e^{(\ell)}_{\overline{\alpha},k}=0,
  \nn\\
\therefore\quad
&C_{\alpha,\pm}
  =\frac{1}{\eta_{\pm}}\bigg[\lambda_{\alpha,\mp}e^{(\ell)}_{\alpha,k}
  +\ii (v^{\prime}_{\alpha}-\ii \Gamma_{\alpha\overline{\alpha}})e^{(\ell)}_{\overline{\alpha},k}\bigg].
\end{eqnarray}
Then, by inserting these into Eqs.~\eqref{eq:Sch-eq_1-state_GDQD_e_4} 
and \eqref{eq:Sch-eq_1-state_GDQD_e_5}, we obtain
\begin{eqnarray}
e_{\alpha}(t)
&=e^{(\ell)}_{\alpha,k}\ee^{-\ii k t}
  \!+\!\sum_{s}\frac{1}{\eta_{s}}
  \bigg[\lambda_{\alpha,\overline{s}}e^{(\ell)}_{\alpha,k}
  \!+\!\ii (v^{\prime}_{\alpha}\!-\!\ii \Gamma_{\alpha\overline{\alpha}})
  e^{(\ell)}_{\overline{\alpha},k}\bigg]
  \ee^{-\ii (\overline{\epsilon}_{{\rm d}}-\ii \overline{\Gamma}+\ii \eta_{s}/2)t}.
\end{eqnarray}
{\bf Step 3)} Prosess  (i), (ii)$\to$(iii):
By using the matching condition of $g_{m}(x,t)$ 
in Eq.~\eqref{eq:Sch-eq_1-state_GDQD_g_3}, 
we have
\begin{eqnarray}
g_{m}(0+,t)
&=\frac{1}{\sqrt{2\pi}}\delta_{m,\ell}\ee^{-\ii kt}
  -\ii \tilde{\theta}(t)\sum_{\alpha}v_{m\alpha}e_{\alpha}(t).
\end{eqnarray}
Here, in order to extend the relation to the region $t\leq 0$, 
we used the step function $\tilde{\theta}(t)$
defined in Eq.~\eqref{eq:step-func}.
Through the translation invariance in Eq.~\eqref{eq:Sch-eq_1-state_GDQD_g_2}, 
we obtain the wave function $g_{m}(x,t)$ for $x>0$ and $t>0$ as
\begin{eqnarray}
g_{m}(x,t)
&=g_{m}(0+,t-x)
 \nn\\
&=\frac{1}{\sqrt{2\pi}}\delta_{m,\ell}\ee^{\ii k(x-t)}
  -\ii \tilde{\theta}(t-x)\sum_{\alpha}v_{m\alpha}e_{\alpha}(t-x).
\end{eqnarray}
The wave functions for $\eta=0$
are obtained in a similar way by using the general solution for $\eta=0$
in Eqs.~\eqref{eq:Sch-eq_1-state_GDQD_e_3}.
\begin{flushright}
$\square$
\end{flushright}

We find that the second terms of the wave functions 
$g^{(\ell)}_{m,k}(x,t)$ and $e^{(\ell)}_{\alpha,k}(t)$ in 
Eqs.~\eqref{eq:1-scattering-state_GDQD_g}, \eqref{eq:1-scattering-state_GDQD_e},
\eqref{eq:1-scattering-state_GDQD_g_ep} and \eqref{eq:1-scattering-state_GDQD_e_ep}
contain the terms with the time-dependent factor
$\ee^{-\ii (\overline{\epsilon}_{{\rm d}}-\ii \overline{\Gamma}+\ii \eta_{s}/2)t}$,
which decays exponentially with the inverse relaxation time $\overline{\Gamma}-{\rm Re}(\eta_{s}/2)$.
The purely exponential decay is due to the unbounded linear dispersion relation~\cite{Gurvitz_15PhysScr};
if there were the lower limit of the energy dispersion,
we would have found deviations from the exponential decay.
We also find that the time-dependent factor corresponds to 
the one-body resonant state, which is explicitly given in \ref{sec:1-ResonantStates}.
The wave functions for $\eta=0$ in Eqs.~\eqref{eq:1-scattering-state_GDQD_g_ep} and 
\eqref{eq:1-scattering-state_GDQD_e_ep} include an exponential function
multiplied by a term linear in time $t$.
We remark that they are reproduced by taking the limit $\eta\to 0$ of 
Eqs.~\eqref{eq:1-scattering-state_GDQD_g} and \eqref{eq:1-scattering-state_GDQD_e}.

In the long-time limit $t\to\infty$,
the wave functions in Eqs.~\eqref{eq:1-scattering-state_GDQD_g}
and \eqref{eq:1-scattering-state_GDQD_e} converge to
the stationary scattering eigenstates~\cite{Nishino-Hatano-Ordonez_16JPC},
\numparts
\begin{eqnarray}
\fl
&\lim_{t\to\infty}\ee^{\ii kt}g^{(\ell)}_{m,k}(x,t)
 =\frac{1}{\sqrt{2\pi}}\bigg[\delta_{m,\ell}
  -\ii \sqrt{2\pi}\theta(x)\sum_{\alpha}v_{m\alpha}e^{(\ell)}_{\alpha,k}
  \bigg]\ee^{\ii kx}
 =: g^{(\ell)}_{m,k}(x),
 \label{eq:1-scattering-eigenstate_GDQD_g}
 \\
\fl
&\lim_{t\to\infty}\ee^{\ii kt}e^{(\ell)}_{\alpha,k}(t)=e^{(\ell)}_{\alpha,k},
 \label{eq:1-scattering-eigenstate_GDQD_e}
\end{eqnarray}
\endnumparts
where $e^{(\ell)}_{\alpha,k}$ is defined in Eq.~\eqref{eq:1-eley_GDQD}.
We remark that
the wave functions $g^{(\ell)}_{m,k}(x)$ and $e^{(\ell)}_{\alpha,k}$ have 
resonance poles at $k=\overline{\epsilon}_{{\rm d}}-\ii \overline{\Gamma}\pm\ii \eta/2$ on the complex $k$ plane.
The imaginary parts of the resonance poles correspond to the relaxation time of 
the time-evolving scattering states.

\mathversion{bold}
\subsection{Two-electron scattering states}
\mathversion{normal}
\label{sec:2-time-evolving-scattering-states}

Next we consider the two-electron case.
In a way similar to the one-electron case,
we derive several relations among the wave functions 
from the set of Schr\"odinger equations~\eqref{eq:Sch-eq_2-state_GDQD_g},
\eqref{eq:Sch-eq_2-state_GDQD_e} and \eqref{eq:Sch-eq_2-state_GDQD_f}.
We consider the case of $\eta\neq 0$ 
since the wave functions for $\eta=0$ are obtained by taking the limit $\eta\to 0$.
First, we find that the wave function $g_{m_{1}m_{2}}(x_{1},x_{2},t)$ is discontinuous 
both at $x_{1}=0$ and $x_{2}=0$. 
In each quadrant of the $(x_{1},x_{2})$-plane,
the general solution $g_{m_{1}m_{2}}(x_{1},x_{2},t)$ is given by
an arbitrary function $F(x_{1}-t,x_{2}-t)$ of the two variables $x_{1}-t$ and $x_{2}-t$.
In other words, the wave function $g_{m_{1}m_{2}}(x_{1},x_{2},t)$ has the following translation invariance:
\begin{eqnarray}
\label{eq:2-state_GDQD_g-invariance}
&g_{m_{1}m_{2}}(x_{1}+a,x_{2}+a,t+a)
  \nn\\
&=g_{m_{1}m_{2}}(x_{1},x_{2},t)\quad
 \mbox{ if } x_{i}(x_{i}+a)>0 \mbox{ for } i=1, 2.
\end{eqnarray}
The matching conditions of $g_{m_{1}m_{2}}(x_{1},x_{2}, t)$ 
at $x_{1}=0$ and $x_{2}=0$ are respectively given by
\begin{eqnarray}
\label{eq:2-state_GDQD_g-matching}
&g_{m_{1}m_{2}}(x,0+,t)-g_{m_{1}m_{2}}(x,0-,t)
  +\ii \sum_{\alpha}v_{m_{2}\alpha}e_{m_{1},\alpha}(x,t)=0,
 \nn\\
&g_{m_{1}m_{2}}(0+,x,t)-g_{m_{1}m_{2}}(0-,x,t)
  -\ii \sum_{\alpha}v_{m_{1}\alpha}e_{m_{2},\alpha}(x,t)=0,
\end{eqnarray}
as was in Eq.~\eqref{eq:Sch-eq_1-state_GDQD_g_3}.
We note that the two relations are consistent with the anti-symmetries
in Eqs.~\eqref{eq:2-state_GDQD_anti-sym}.
Similarly to Eq.~\eqref{eq:Sch-eq_1-state_GDQD_g_4},
we assume the values of $g_{m_{1}m_{2}}(x_{1},x_{2},t)$ at $x_{1}=0$ and $x_{2}=0$ as follows:
\begin{eqnarray}
\label{eq:2-state_GDQD_g-average}
&g_{m_{1}m_{2}}(0,x,t)=\frac{1}{2}\big(g_{m_{1}m_{2}}(0+,x,t)+g_{m_{1}m_{2}}(0-,x,t)\big),
 \nn\\
&g_{m_{1}m_{2}}(x,0,t)=\frac{1}{2}\big(g_{m_{1}m_{2}}(x,0+,t)+g_{m_{1}m_{2}}(x,0-,t)\big).
\end{eqnarray}

Next, we find that the wave function $e_{m,\alpha}(x,t)$ is also discontinuous at $x=0$,
and its matching condition at $x=0$ is given by
\begin{eqnarray}
\label{eq:2-state_GDQD_e-matching}
&e_{m,\alpha}(0+,t)-e_{m,\alpha}(0-,t)
 -\ii v_{m\overline{\alpha}}f_{\alpha\overline{\alpha}}(t)=0.
\end{eqnarray}
We assume the value of $e_{m,\alpha}(x,t)$ at $x=0$ as follows:
\begin{eqnarray}
\label{eq:2-state_GDQD_e-average}
&e_{m,\alpha}(0,t)=\frac{1}{2}\big(e_{m,\alpha}(0+,t)+e_{m,\alpha}(0-,t)\big).
\end{eqnarray}
By applying Eqs.~\eqref{eq:2-state_GDQD_g-matching} and \eqref{eq:2-state_GDQD_g-average}
to Eq.~\eqref{eq:Sch-eq_2-state_GDQD_e} for $x\neq 0$, we have
the coupled partial-differential equations for $e_{m,\alpha}(x,t)$ and $e_{m,\overline{\alpha}}(x,t)$ as
\begin{eqnarray}
\fl
&\big(\ii (\partial_{t}+\partial_{x})-\epsilon_{{\rm d}\alpha}+\ii \Gamma_{\alpha\alpha}\big)
 e_{m,\alpha}(x,t)
 -\big(v^{\prime}_{\alpha}-\ii \Gamma_{\alpha\overline{\alpha}}\big)e_{m,\overline{\alpha}}(x,t)
 =\sum_{n}v^{\ast}_{n\alpha}g_{mn}(x,0-,t).
\end{eqnarray}
Through the change of variables $z=(x+t)/2$ and $z_{1}=x-t$,
the partial-differential equation is rewritten as
\begin{eqnarray}
&\big(\ii \partial_{z}-\epsilon_{{\rm d}\alpha}+\ii \Gamma_{\alpha\alpha}\big)
 e_{m,\alpha}(x,t)
 -\big(v^{\prime}_{\alpha}-\ii \Gamma_{\alpha\overline{\alpha}}\big)e_{m,\overline{\alpha}}(x,t)
 \nn\\
&=\sum_{n}v^{\ast}_{n\alpha}g_{mn}\Big(z+\frac{z_{1}}{2},0-,z-\frac{z_{1}}{2}\Big).
 \label{eq:2-state_GDQD_e-PDE_1}
\end{eqnarray}
By introducing an auxiliary function
$e_{m,\alpha}(x,t)=\ee^{-\ii (\epsilon_{{\rm d}\alpha}-\ii \Gamma_{\alpha\alpha})z}
\tilde{e}_{m,\alpha}(x,t)$, we obtain 
\begin{eqnarray}
&\ii \partial_{z}\tilde{e}_{m,\alpha}(x,t)
 =\big(v^{\prime}_{\alpha}-\ii \Gamma_{\alpha\overline{\alpha}}\big)
 \ee^{\ii (\Delta\epsilon_{{\rm d}\alpha}-\ii \Delta\Gamma_{\alpha})z}
 \tilde{e}_{m,\overline{\alpha}}(x,t)
 \nn\\
&\hspace{75pt}
 +\sum_{n}v^{\ast}_{n\alpha}
 \ee^{\ii (\epsilon_{{\rm d}\alpha}-\ii \Gamma_{\alpha\alpha})z}
 g_{mn}\Big(z+\frac{z_{1}}{2},0-,z-\frac{z_{1}}{2}\Big),
 \label{eq:2-state_GDQD_e-PDE_2}
\end{eqnarray}
which is transformed to a second-order differential equation
for $\tilde{e}_{m,\alpha}(x,t)$ for each $\alpha$ as
\begin{eqnarray}
&\partial^{2}_{z}\tilde{e}_{m,\alpha}(x,t)
 -\ii (\Delta\epsilon_{{\rm d}\alpha}\!-\!\ii \Delta\Gamma_{\alpha})\partial_{z}\tilde{e}_{m,\alpha}(x,t)
 \nn\\
&+(v^{\prime}_{\alpha}-\ii \Gamma_{\alpha\overline{\alpha}})
  (v^{\prime}_{\overline{\alpha}}-\ii \Gamma_{\overline{\alpha}\alpha})
  \tilde{e}_{m,\alpha}(x,t)
 =G_{m,\alpha}(z)
 \label{eq:2-state_GDQD_e-PDE_3}
\end{eqnarray}
with the inhomogeneous term 
\begin{eqnarray}
\label{eq:2-state_GDQD_e-G}
G_{m,\alpha}(z)
&=\!-\!\sum_{n}\ee^{\ii (\epsilon_{{\rm d}\alpha}-\ii \Gamma_{\alpha\alpha})z}
 \bigg[v^{\ast}_{n\alpha}\big(\ii \partial_{z}\!-\!\epsilon_{{\rm d}\overline{\alpha}}
  \!+\!\ii \Gamma_{\overline{\alpha}\,\overline{\alpha}})
 +v^{\ast}_{n\overline{\alpha}}(v^{\prime}_{\alpha}\!-\!\ii \Gamma_{\alpha\overline{\alpha}})\bigg]
 \nn\\
&\quad\times
 g_{mn}\Big(z\!+\!\frac{z_{1}}{2},0-,z\!-\!\frac{z_{1}}{2}\Big).
\end{eqnarray}
The general solution of Eq.~\eqref{eq:2-state_GDQD_e-PDE_3} is given by
\begin{eqnarray}
\fl
&\tilde{e}_{m,\alpha}(x,t)
 =\sum_{s}C_{m,\alpha,s}(x\!-\! t)\ee^{\lambda_{\alpha,s}(x+t)/2}
 \nn\\
\fl
&\hspace{57pt}
 +\sum_{s}\frac{1}{\lambda_{\alpha,s}\!-\!\lambda_{\alpha,\overline{s}}}
  \int^{(x+t)/2}_{-\infty}
  \hspace{-25pt}dz^{\prime}\,\ee^{\lambda_{\alpha,s}((x+t)/2-z^{\prime})}G_{m,\alpha}(z^{\prime}),
 \nn\\
\fl
\therefore\quad
&e_{m,\alpha}(x,t)
 =\sum_{s}C_{m,\alpha,s}(x\!-\! t)
  \ee^{-\ii (\epsilon_{{\rm d}\alpha}-\ii \Gamma_{\alpha\alpha}+\ii \lambda_{\alpha,s})(x+t)/2}
 \nn\\
\fl
&\hspace{57pt}
 +\sum_{s}\frac{1}{\lambda_{\alpha,s}\!-\!\lambda_{\alpha,\overline{s}}}
   \int_{-\infty}^{(x+t)/2}
   \hspace{-25pt}dz^{\prime}\,\ee^{-\ii (\epsilon_{{\rm d}\alpha}-\ii \Gamma_{\alpha\alpha})(x+t)/2
   +\lambda_{\alpha,s}((x+t)/2-z^{\prime})}G_{m,\alpha}(z^{\prime}),
  \label{eq:2-state_GDQD_e-general}
\end{eqnarray}
where $C_{m,\alpha,s}(x-t)$ for $s=\pm$ are arbitrary functions of the variable $x-t$.

Finally, by using Eqs.~\eqref{eq:2-state_GDQD_e-matching} and \eqref{eq:2-state_GDQD_e-average},
the differential equation~\eqref{eq:Sch-eq_2-state_GDQD_f}
for the double-occupancy wave function $f_{\alpha\overline{\alpha}}(t)$ is rewritten as
\begin{eqnarray}
\label{eq:2-state_GDQD_f-PDE_1}
&(\ii \partial_{t}-2\overline{\epsilon}_{\rm d}-U^{\prime}+2\ii \overline{\Gamma})f_{\alpha\overline{\alpha}}(t)
 =\sum_{m, \beta}
  (-)^{\alpha+\beta}v^{\ast}_{m\beta}e_{m,\overline{\beta}}(0-,t).
\end{eqnarray}
By setting $f_{\alpha\overline{\alpha}}(t)=\ee^{-\ii (2\overline{\epsilon}_{\rm d}+U^{\prime}-2\ii \overline{\Gamma})t}
\tilde{f}_{\alpha\overline{\alpha}}(t)$, we have
\begin{eqnarray}
\label{eq:2-state_GDQD_f-PDE_2}
&\partial_{t}\tilde{f}_{\alpha\overline{\alpha}}(t)
 =-\ii \ee^{\ii (2\overline{\epsilon}_{\rm d}+U^{\prime}-2\ii \overline{\Gamma})t}
  \sum_{m, \beta}
  (-)^{\alpha+\beta}v^{\ast}_{m\beta}e_{m,\overline{\beta}}(0-,t),
\end{eqnarray}
which is easily integrated in the form
\begin{eqnarray}
\fl
&\tilde{f}_{\alpha\overline{\alpha}}(t)
 =C_{\alpha}
  -\ii \int^{t}_{0}\hspace{-4pt}dt^{\prime}\,
  \ee^{\ii (2\overline{\epsilon}_{\rm d}+U^{\prime}-2\ii \overline{\Gamma})t^{\prime}}
  \sum_{m, \beta}
  (-)^{\alpha+\beta}v^{\ast}_{m\beta}e_{m,\overline{\beta}}(0-,t^{\prime}),
  \nn\\
\fl
\therefore\quad
&f_{\alpha\overline{\alpha}}(t)
 =C_{\alpha}\ee^{-\ii (2\overline{\epsilon}_{\rm d}+U^{\prime}-2\ii \overline{\Gamma})t}
 \nn\\
\fl
&\hspace{42pt}
  -\ii \int^{t}_{0}\hspace{-4pt}dt^{\prime}\,
  \ee^{\ii (2\overline{\epsilon}_{\rm d}+U^{\prime}-2\ii \overline{\Gamma})(t^{\prime}-t)}
  \sum_{m, \beta}
  (-)^{\alpha+\beta}v^{\ast}_{m\beta}e_{m,\overline{\beta}}(0-,t^{\prime}).
  \label{eq:2-state_GDQD_f-solution}
\end{eqnarray}
Here $C_{\alpha}$ for $\alpha=1, 2$ are integration constants. 

\begin{figure}[t]
\begin{center}
{
\begin{picture}(360,150)(0,0)
 \put(0,0){\includegraphics[width=350pt]{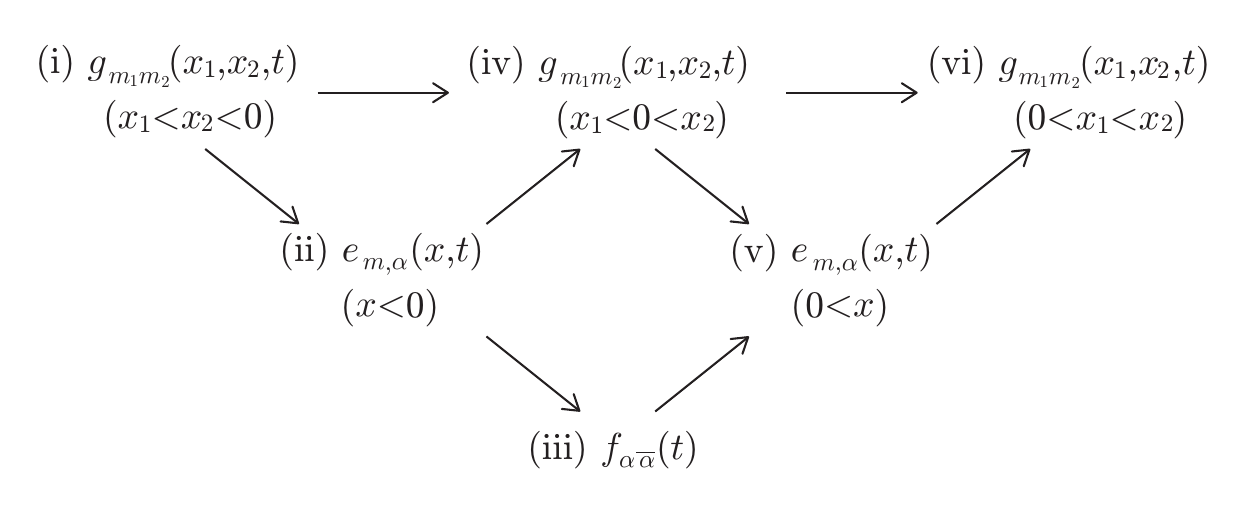}}
\end{picture}
}
\end{center}
\caption{The flow chart of the construction of time-evolving two-electron states.}
\label{fig:flow-chart_2-elec}
\end{figure}

Now we propose a systematic construction of 
the time-evolving two-electron states under arbitrary initial conditions.
The construction is given by the flow chart in Fig.~\ref{fig:flow-chart_2-elec}:
\begin{description}
\itemsep=0pt
\item[Step 1)]
The wave function $g_{m_{1}m_{2}}(x_{1},x_{2},t)$ for $x_{1}<x_{2}<0$ 
in (i) of Fig.~\ref{fig:flow-chart_2-elec}
is obtained by the translation invariance in Eq.~\eqref{eq:2-state_GDQD_g-invariance}
with the wave function $g_{m_{1}m_{2}}(x_{1},x_{2},0)$ of given initial states.
\item[Step 2)] 
In the process (i)$\to$(ii) in Fig.~\ref{fig:flow-chart_2-elec}, 
we use the general solution $e_{m,\alpha}(x,t)$ in Eq.~\eqref{eq:2-state_GDQD_e-general}.
The arbitrary function $C_{m,\alpha,s}(x-t)$
is determined by the initial condition for $e_{m,\alpha}(x,t)$.
\item[Step 3)]
In the process (ii)$\to$(iii) in Fig.~\ref{fig:flow-chart_2-elec}, 
we use the general solution $f_{\alpha\overline{\alpha}}(t)$
in Eq.~\eqref{eq:2-state_GDQD_f-solution}.
The integration constant $C_{\alpha}$
is determined by the initial condition for $f_{\alpha\overline{\alpha}}(t)$.
\item[Step 4)]
In the process (i), (ii)$\to$(iv) in Fig.~\ref{fig:flow-chart_2-elec}, 
we use the matching conditions $g_{m_{1}m_{2}}(x_{1},x_{2},t)$
in Eqs.~\eqref{eq:2-state_GDQD_g-matching}.
\item[Step 5)]
In the process (iii), (iv)$\to$(v) in Fig.~\ref{fig:flow-chart_2-elec}, 
we use the general solution $e_{m,\alpha}(x,t)$
in Eq.~\eqref{eq:2-state_GDQD_e-general} again.
The arbitrary function $C_{m,\alpha,s}(x-t)$
is determined by the initial condition for $e_{m,\alpha}(x,t)$
and the matching condition for $e_{m,\alpha}(x,t)$ in Eq.~\eqref{eq:2-state_GDQD_e-matching}.
\item[Step 6)]
In the process (iv), (v)$\to$(vi) in Fig.~\ref{fig:flow-chart_2-elec}, 
we use the matching conditions for $g_{m_{1}m_{2}}(x_{1},x_{2},t)$
in Eqs.~\eqref{eq:2-state_GDQD_g-matching} again.
\item[Step 7)]
Through the anti-symmetrization in the variables $x_{1}$ and $x_{2}$, 
we obtain the wave function $g_{m_{1}m_{2}}(x_{1},x_{2},t)$
in the case $x_{2}<x_{1}$.
\end{description}

Let us demonstrate the construction of the time-evolving two-electron scattering states 
under the initial conditions
\begin{eqnarray}
\fl
&g_{m_{1}m_{2}}(x_{1},x_{2},0)
  =\frac{1}{2\pi}\sum_{P}\sgn(P)
    \delta_{m_{1},\ell_{P_{1}}}\delta_{m_{2},\ell_{P_{2}}}
    \ee^{\ii (k_{P_{1}}x_{1}+k_{P_{2}}x_{2})}\, \mbox{ for } m_{1}, m_{2}=1, 2,
   \nn\\
 \fl
&e_{m,\alpha}(x,0)=0,\quad
  f_{\alpha\overline{\alpha}}(0)=0\, \mbox{ for } m=1, 2 \mbox{ and } \alpha=1, 2,
\label{eq:2-initial-state_GDQD}
\end{eqnarray}
where $P=(P_{1},P_{2})$ is a permutation of $(1,2)$.
The initial state stands for
the two-electron plane-wave state with wave numbers $k_{1}$ and $k_{2}$
on the entire leads $\ell_{1}$ and $\ell_{2}$, respectively,
and the absence of electrons on the quantum dots.

\begin{prop}
For $\eta\neq 0$,
the solution of the set of Schr\"odinger equations~\eqref{eq:Sch-eq_2-state_GDQD_g},
\eqref{eq:Sch-eq_2-state_GDQD_e} and \eqref{eq:Sch-eq_2-state_GDQD_f}
for the initial conditions~\eqref{eq:2-initial-state_GDQD} is given by
\numparts
\begin{eqnarray}
\label{eq:2-scattering-state_GDQD_g}
\fl
&g^{(\ell_{1}\ell_{2})}_{m_{1}m_{2},k_{1}k_{2}}(x_{1},x_{2},t)
 =\sum_{P}\sgn(P)
 g^{(\ell_{P_{1}})}_{m_{1},k_{P_{1}}}(x_{1},t)
 g^{(\ell_{P_{2}})}_{m_{2},k_{P_{2}}}(x_{2},t)
 \nn\\
\fl
&-\ii \!\sum_{Q, \alpha, s}\!\sgn(Q)v_{m_{Q_{1}}\alpha}y_{m_{Q_{2}}\alpha,s}
  X^{(\ell_{1}\ell_{2})}_{\alpha,k_{1}k_{2}}(x_{Q_{2}}\!\!-t) 
 \ee^{-\ii (\overline{\epsilon}_{\rm d}-\ii \overline{\Gamma}+\ii \eta_{s}/2)x_{Q_{2}Q_{1}}}
 \theta(x_{Q_{2}Q_{1}})\theta(x_{Q_{1}}),
 \\
\label{eq:2-scattering-state_GDQD_e}
\fl
&e^{(\ell_{1}\ell_{2})}_{m\alpha,k_{1}k_{2}}(x,t)
 \nn\\
\fl
&=\sum_{P}\sgn(P)g^{(\ell_{P_{1}})}_{m,k_{P_{1}}}(x,t)
  e^{(\ell_{P_{2}})}_{\alpha,k_{P_{2}}}(t)
  -\sum_{s}y_{m\alpha,s}X^{(\ell_{1}\ell_{2})}_{\alpha,k_{1}k_{2}}(x-t)
  \ee^{-\ii (\overline{\epsilon}_{\rm d}-\ii \overline{\Gamma}+\ii \eta_{s}/2)x}\theta(x),
 \\
\label{eq:2-scattering-state_GDQD_f}
\fl
&f^{(\ell_{1}\ell_{2})}_{\alpha\overline{\alpha},k_{1}k_{2}}(t)
 =\sum_{P}\sgn(P)
 e^{(\ell_{P_{1}})}_{\alpha,k_{P_{1}}}(t)
 e^{(\ell_{P_{2}})}_{\overline{\alpha},k_{P_{2}}}(t)
 +X^{(\ell_{1}\ell_{2})}_{\alpha,k_{1}k_{2}}(-t),
\end{eqnarray}
\endnumparts
where $P$ and $Q$ are permutations of $(1,2)$, $x_{12}=x_{1}-x_{2}$ and we introduced
\begin{eqnarray}
\label{eq:2-state_GDQD_X}
\fl
&X^{(\ell_{1}\ell_{2})}_{\alpha,k_{1}k_{2}}(-t)
 =\sum_{P}\sgn(P)
  \Big[\frac{1}{2}Z^{(\ell_{1}\ell_{2})}_{\alpha,k_{1}k_{2}}
  \big(\ee^{-\ii (k_{1}+k_{2})t}
  -\ee^{-\ii (2\overline{\epsilon}_{\rm d}+U^{\prime}-2\ii \overline{\Gamma})t}\big)
  \nn\\
\fl
&\hspace{75pt}\quad
 -e^{(\ell_{P_{1}})}_{\alpha,k_{P_{1}}}
  e^{(\ell_{P_{2}})}_{\overline{\alpha},k_{P_{2}}}
  \big(\ee^{-2\ii (\overline{\epsilon}_{\rm d}-\ii \overline{\Gamma})t}
  -\ee^{-\ii (2\overline{\epsilon}_{\rm d}+U^{\prime}-2\ii \overline{\Gamma})t}\big)
  \nn\\
\fl
&\hspace{75pt}
 +\frac{1}{\sqrt{2\pi}}\sum_{\beta, s}(-)^{\alpha+\beta}
  \frac{\ii \tilde{y}_{\ell_{P_{1}}\beta,s}U^{\prime}}
  {(k_{P_{1}}\!-\!\overline{\epsilon}_{\rm d}\!+\!\ii \overline{\Gamma}\!-\!\ii \eta_{s}/2)
  (k_{P_{1}}\!-\!\overline{\epsilon}_{\rm d}\!-\! U^{\prime}
  \!+\!\ii \overline{\Gamma}\!-\!\ii \eta_{s}/2)}
  e^{(\ell_{P_{2}})}_{\overline{\beta},k_{P_{2}}}
  \nn\\
\fl
&\hspace{75pt}
 \quad\times
  \big(\ee^{-\ii (2\overline{\epsilon}_{\rm d}+U^{\prime}-2\ii \overline{\Gamma})t}
  - \ee^{-\ii (k_{P_{1}}+\overline{\epsilon}_{\rm d}-\ii \overline{\Gamma}-\ii \eta_{s}/2)t}\big)
 \Big]\tilde{\theta}(t),
 \nn\\
\fl
&Z^{(\ell_{1}\ell_{2})}_{\alpha,k_{1}k_{2}}
 =\frac{U^{\prime}}
  {k_{1}\!+\! k_{2}\!-\! 2\overline{\epsilon}_{\rm d}\!-\! U^{\prime}\!+\! 2\ii \overline{\Gamma}}
  \big(e^{(\ell_{1})}_{\alpha,k_{1}}e^{(\ell_{2})}_{\overline{\alpha},k_{2}}
  -e^{(\ell_{2})}_{\alpha,k_{2}}e^{(\ell_{1})}_{\overline{\alpha},k_{1}}\big).
\end{eqnarray}
\end{prop}
\noindent
{\it Proof.}
We follow the seven steps above.

\noindent
{\bf Step 1)} 
Since we have $x_{1}-t<x_{2}-t<0$ for $t>0$ and $x_{1}<x_{2}<0$ in this step, 
the wave function $g_{m_{1}m_{2}}(x_{1},x_{2},t)$ is obtained 
through the invariance of the translation $(x_{1}, x_{2}, t)\mapsto (x_{1}-t, x_{2}-t, 0)$ 
from the initial state $g_{m_{1}m_{2}}(x_{1}-t,x_{2}-t,0)$ as
\begin{eqnarray}
g_{m_{1}m_{2}}(x_{1},x_{2},t)
&=g_{m_{1}m_{2}}(x_{1}-t,x_{2}-t,0)
  \nn\\
&=\frac{1}{2\pi}\sum_{P}\sgn(P)
    \delta_{m_{1},\ell_{P_{1}}}\delta_{m_{2},\ell_{P_{2}}}
    \ee^{\ii k_{P_{1}}(x_{1}-t)+\ii k_{P_{2}}(x_{2}-t)}.
\end{eqnarray}

\noindent
{\bf Step 2)} Process (i)$\to$(ii): By setting $x_{2}=0-$ in the above equation, we have
\begin{eqnarray}
&g_{m_{1}m_{2}}(x_{1},0-,t)
  =\frac{1}{2\pi}\sum_{P}\sgn(P)
    \delta_{m_{1},\ell_{P_{1}}}\delta_{m_{2},\ell_{P_{2}}}
    \ee^{\ii k_{P_{1}}(x_{1}-t)-\ii k_{P_{2}}t},
\end{eqnarray}
which gives the inhomogeneous term in Eq.~\eqref{eq:2-state_GDQD_e-G} in the form
\begin{eqnarray}
\fl
G_{m,\alpha}(z)
&=-\sum_{n}
 \bigg[v^{\ast}_{n\alpha}\big(k_{P_{2}}-\epsilon_{{\rm d}\overline{\alpha}}
  +\ii \Gamma_{\overline{\alpha}\,\overline{\alpha}})
 +v^{\ast}_{n\overline{\alpha}}(v^{\prime}_{\alpha}-\ii \Gamma_{\alpha\overline{\alpha}})\bigg]
 \nn\\
\fl
&\quad\times
 \frac{1}{2\pi}\sum_{P}\sgn(P)
    \delta_{m,\ell_{P_{1}}}\delta_{n,\ell_{P_{2}}}
    \ee^{\ii (k_{P_{1}}+k_{P_{2}}/2)z_{1}
    +\ii (-k_{P_{2}}+\epsilon_{{\rm d}\alpha}-\ii \Gamma_{\alpha\alpha})z}.
\end{eqnarray}
Then, for $x<0$, we obtain the general solution in Eq.~\eqref{eq:2-state_GDQD_e-general} as
\numparts
\begin{eqnarray}
e_{m,\alpha}(x,t)
&=\sum_{s}C_{m,\alpha,s}(x-t)
  \ee^{\ii (-\epsilon_{{\rm d}\alpha}+\ii \Gamma_{\alpha\alpha}-\ii \lambda_{\alpha,s})(x+t)/2}
  \nn\\
&\quad
  +\sum_{P}\sgn(P)
  g^{(\ell_{P_{1}})}_{m,k_{P_{1}}}(x,t)
  e^{(\ell_{P_{2}})}_{\alpha,k_{P_{2}}}
  \ee^{-\ii k_{P_{2}}t},
  \label{eq:2-state_GDQD_e-general_2_1}
\\
e_{m,\overline{\alpha}}(x,t)
&=\sum_{s}\frac{\ii \lambda_{\alpha,s}}{v^{\prime}_{\alpha}\!-\!\ii \Gamma_{\alpha\overline{\alpha}}}
  C_{m,\alpha,s}(x-t)
  \ee^{\ii (-\epsilon_{{\rm d}\overline{\alpha}}
  +\ii \Gamma_{\overline{\alpha}\overline{\alpha}}-\ii \lambda_{\overline{\alpha},s})(x+t)/2}
 \nn\\
&\quad
 +\sum_{P}\sgn(P)
  g^{(\ell_{P_{1}})}_{m,k_{P_{1}}}(x,t)
  e^{(\ell_{P_{2}})}_{\overline{\alpha},k_{P_{2}}}
  \ee^{-\ii k_{P_{2}}t},
 \label{eq:2-state_GDQD_e-general_2_2}
\end{eqnarray}
\endnumparts
where $e_{m,\overline{\alpha}}(x,t)$ is obtained by using Eq.~\eqref{eq:2-state_GDQD_e-PDE_1}
with the above expression for $e_{m,\alpha}(x,t)$.
Here we used  the wave function $g^{(\ell)}_{m,k}(x,t)$ of the one-electron scattering states
in Eq.~\eqref{eq:1-scattering-state_GDQD_g}.
The initial conditions in Eqs.~\eqref{eq:2-initial-state_GDQD} 
for $e_{m,\alpha}(x,t)$ and $e_{m,\overline{\alpha}}(x,t)$
determine the arbitrary functions
$C_{m,\alpha,s}(x)$ and $C_{m,\alpha,\overline{s}}(x)$ as
\begin{eqnarray}
\fl
&e_{m,\alpha}(x,0+)
 =\sum_{s}C_{m,\alpha,s}(x)
  \ee^{\ii (-\epsilon_{{\rm d}\alpha}+\ii \Gamma_{\alpha\alpha}-\ii \lambda_{\alpha,s})x/2}
  +\sum_{P}\sgn(P)
  g^{(\ell_{P_{1}})}_{m,k_{P_{1}}}(x,0)
  e^{(\ell_{P_{2}})}_{\alpha,k_{P_{2}}}=0,
  \nn\\
\fl
&e_{m,\overline{\alpha}}(x,0+)
 =\sum_{s}
  \frac{\ii \lambda_{\alpha,s}}{v^{\prime}_{\alpha}\!-\!\ii \Gamma_{\alpha\overline{\alpha}}}
  C_{m,\alpha,s}(x)
  \ee^{\ii (-\epsilon_{{\rm d}\overline{\alpha}}
  +\ii \Gamma_{\overline{\alpha}\,\overline{\alpha}}-\ii \lambda_{\overline{\alpha},s})x/2}
 \nn\\
\fl
&\hspace{70pt}
  +\sum_{P}\sgn(P)
  g^{(\ell_{P_{1}})}_{m,k_{P_{1}}}(x,0)
  e^{(\ell_{P_{2}})}_{\overline{\alpha},k_{P_{2}}}=0,
 \nn\\
\fl
\therefore\quad
&C_{m,\alpha,s}(x)
 =\frac{1}{\eta_{s}}
  \sum_{P}\sgn(P)
  g^{(\ell_{P_{1}})}_{m,k_{P_{1}}}(x,0)
  \bigg[\lambda_{\alpha,\overline{s}}e^{(\ell_{P_{2}})}_{\alpha,k_{P_{2}}}
  +\ii (v^{\prime}_{\alpha}-\ii \Gamma_{\alpha\overline{\alpha}})
  e^{(\ell_{P_{2}})}_{\overline{\alpha},k_{P_{2}}}\bigg]
  \nn\\
\fl
&\hspace{60pt}\times
  \ee^{\ii (\epsilon_{{\rm d}\alpha}-\ii \Gamma_{\alpha\alpha}+\ii \lambda_{\alpha,s})x/2}.
 \label{eq:2-state_GDQD_e-C_1}
\end{eqnarray}
By using the translation invariance $g^{(\ell_{P_{1}})}_{m,k_{P_{1}}}(x,t)
=g^{(\ell_{P_{1}})}_{m,k_{P_{1}}}(x-t,0)$ for $t>0$ and $x<0$, we have
\begin{eqnarray}
C_{m,\alpha,s}(x-t)
&=\frac{1}{\eta_{s}}\sum_{P}\sgn(P)
  g^{(\ell_{P_{1}})}_{m,k_{P_{1}}}(x,t)
  \bigg[\lambda_{\alpha,\overline{s}}e^{(\ell_{P_{2}})}_{\alpha,k_{P_{2}}}
  +\ii (v^{\prime}_{\alpha}-\ii \Gamma_{\alpha\overline{\alpha}})
  e^{(\ell_{P_{2}})}_{\overline{\alpha},k_{P_{2}}}\bigg]
  \nn\\
&\quad\times
  \ee^{\ii (\epsilon_{{\rm d}\alpha}-\ii \Gamma_{\alpha\alpha}+\ii \lambda_{\alpha,s})(x-t)/2}.
\end{eqnarray}
By inserting this into Eqs.~\eqref{eq:2-state_GDQD_e-general_2_1}
and \eqref{eq:2-state_GDQD_e-general_2_2}, we obtain
$e_{m,\alpha}(x,t)$ and $e_{m,\overline{\alpha}}(x,t)$ in the same form
\begin{eqnarray}
e_{m,\alpha}(x,t)
&=\sum_{P}\sgn(P)g^{(\ell_{P_{1}})}_{m,k_{P_{1}}}(x,t)
  e^{(\ell_{P_{2}})}_{\alpha,k_{P_{2}}}(t).
 \label{eq:2-state_GDQD_e_1}
\end{eqnarray}
Here $e^{(\ell)}_{\alpha,k}(t)$ is the wave function of
the one-electron scattering state in Eq.~\eqref{eq:1-scattering-state_GDQD_e}.

\noindent
{\bf Step 3)} Process (ii)$\to$(iii):
We insert Eq.~\eqref{eq:2-state_GDQD_e_1}
into the solution in Eq.~\eqref{eq:2-state_GDQD_f-solution}
for $f_{\alpha\overline{\alpha}}(t)$.
Because of the initial condition $f_{\alpha\overline{\alpha}}(0)=0$, we find $C=0$, obtaining
\begin{eqnarray}
\fl
f_{\alpha\overline{\alpha}}(t)
 =\sum_{P}\sgn(P)
  \Big[\frac{k_{P_{1}}\!+\! k_{P_{2}}\!-\! 2\overline{\epsilon}_{\rm d}\!+\! 2\ii \overline{\Gamma}}
  {k_{P_{1}}\!+\! k_{P_{2}}\!-\! 2\overline{\epsilon}_{\rm d}\!-\! U^{\prime}\!+\! 2\ii \overline{\Gamma}}
  e^{(\ell_{P_{1}})}_{\alpha,k_{P_{1}}}e^{(\ell_{P_{2}})}_{\overline{\alpha},k_{P_{2}}}
  \big(\ee^{-\ii (k_{P_{1}}+k_{P_{2}})t}\!-\!
  \ee^{-\ii (2\overline{\epsilon}_{\rm d}+U^{\prime}-2\ii \overline{\Gamma})t}\big)
 \nn\\
\fl
 \hspace{50pt}+\frac{1}{\sqrt{2\pi}}\sum_{\beta}(-)^{\alpha+\beta}e^{(\ell_{P_{2}})}_{\beta,k_{P_{2}}}
  \sum_{s}\frac{\ii \tilde{y}_{\ell_{P_{1}}\overline{\beta},\overline{s}}}
  {k_{P_{1}}\!-\!\overline{\epsilon}_{\rm d}\!-\! U^{\prime}\!+\!\ii \overline{\Gamma}\!-\!\ii \eta_{\overline{s}}/2}
  \nn\\
\fl
 \hspace{50pt}\quad\times
  \big(\ee^{-\ii (k_{P_{1}}+\overline{\epsilon}_{\rm d}-\ii \overline{\Gamma}+\ii \eta_{s}/2)t}
  -\ee^{-\ii (2\overline{\epsilon}_{\rm d}+U^{\prime}-2\ii \overline{\Gamma})t}\big)\Big].
\end{eqnarray}
After some calculations, we arrive at the expression
\begin{eqnarray}
&f_{\alpha\overline{\alpha}}(t)
   =\sum_{P}\sgn(P)
 e^{(\ell_{P_{1}})}_{\alpha,k_{P_{1}}}(t)
 e^{(\ell_{P_{2}})}_{\overline{\alpha},k_{P_{2}}}(t)
 +X^{(\ell_{1}\ell_{2})}_{\alpha,k_{1}k_{2}}(-t),
 \label{eq:2-state_GDQD_f_1}
\end{eqnarray}
with $X^{(\ell_{1}\ell_{2})}_{\alpha,k_{1}k_{2}}(-t)$
defined in Eqs.~\eqref{eq:2-state_GDQD_X}.

\noindent
{\bf Step 4)}  Process (i), (ii)$\to$(iv):
The first matching condition in Eqs.~\eqref{eq:2-state_GDQD_g-matching} gives
\begin{eqnarray}
 g_{m_{1}m_{2}}(x,0+,t)
&=\sum_{P}\sgn(P)g^{(\ell_{P_{1}})}_{m_{1},k_{P_{1}}}(x,t)g^{(\ell_{P_{2}})}_{m_{2},k_{P_{1}}}(0+,t).
\end{eqnarray}
Since $x_{1}-x_{2}<0$ for $x_{1}<0<x_{2}$ in this step, we have
\begin{eqnarray}
g_{m_{1}m_{2}}(x_{1},x_{2},t)
&=g_{m_{1}m_{2}}(x_{1}-x_{2},0+,t-x_{2})
 \nn\\
&=\sum_{P}\sgn(P)g^{(\ell_{P_{1}})}_{m_{1},k_{P_{1}}}(x_{1},t)
 g^{(\ell_{P_{2}})}_{m_{2},k_{P_{1}}}(x_{2},t)
 \label{eq:2-state_GDQD_g_2}
\end{eqnarray}
from the translation invariance of the wave function $g_{m_{1}m_{2}}(x_{1},x_{2},t)$ 
in Eq.~\eqref{eq:2-state_GDQD_g-invariance}.

\noindent
{\bf Step 5)}  Process (iii), (iv)$\to$(v):
By setteing $x_{1}=0-$ in Eq.~\eqref{eq:2-state_GDQD_g_2}, we have
\begin{eqnarray}
&g_{m_{1}m_{2}}(0-,x,t)
 =\sum_{P}\sgn(P)g^{(\ell_{P_{1}})}_{m_{1},k_{P_{1}}}(0-,t)
 g^{(\ell_{P_{2}})}_{m_{2},k_{P_{1}}}(x,t),
\end{eqnarray}
which gives the inhomogeneous term in Eq.~\eqref{eq:2-state_GDQD_e-G} as 
\begin{eqnarray}
\fl
&G_{n,\alpha}(z)
 =\sum_{m}
 \bigg[v^{\ast}_{m\alpha}\big(k_{P_{1}}\!-\!\epsilon_{{\rm d}\overline{\alpha}}
  \!+\!\ii \Gamma_{\overline{\alpha}\,\overline{\alpha}})
 +v^{\ast}_{m\overline{\alpha}}(v^{\prime}_{\alpha}\!-\!\ii \Gamma_{\alpha\overline{\alpha}})\bigg]
 \nn\\
\fl
&\hspace{50pt}\times
 \frac{1}{\sqrt{2\pi}}\sum_{P}\sgn(P)\delta_{m\ell_{P_{1}}}\ee^{\ii k_{P_{1}}z_{1}/2}
 \ee^{\ii (-k_{P_{1}}+\epsilon_{{\rm d}\alpha}-\ii \Gamma_{\alpha\alpha})z}
 \nn\\
\fl
&\hspace{50pt}\times
 \Big[\frac{1}{\sqrt{2\pi}}\delta_{n\ell_{P_{2}}}\ee^{\ii k_{P_{2}}z_{1}}
 +\sum_{\beta, s}e^{(\ell_{P_{2}})}_{\beta,k_{P_{2}}}\, y_{n\overline{\beta},\overline{s}}
  \big(\ee^{\ii k_{P_{2}}z_{1}}
  \!-\!\ee^{\ii (\overline{\epsilon}_{\rm d}-\ii \overline{\Gamma}+\ii \eta_{s}/2)z_{1}}\big)
  \theta(-z_{1})\Big].
\end{eqnarray}
Then, for $x>0$, we obtain the general solution in Eq.~\eqref{eq:2-state_GDQD_e-general} as
\numparts
\begin{eqnarray}
 e_{n,\alpha}(x,t)
&=\sum_{s}C_{n,\alpha,s}(x-t)
  \ee^{-\ii (\overline{\epsilon}_{{\rm d}}-\ii \overline{\Gamma}+\frac{\ii }{2}\eta_{s})(x+t)/2}
 \nn\\
&\quad
 -\sum_{P}\sgn(P)e^{(\ell_{P_{1}})}_{\alpha,k_{P_{1}}}
  \ee^{-\ii k_{P_{1}}t}g^{(\ell_{P_{2}})}_{n,k_{P_{2}}}(x,t),
  \label{eq:2-state_GDQD_e-general_3_1}
  \\
 e_{n,\overline{\alpha}}(x,t)
&=\sum_{s}\frac{\ii \lambda_{\alpha,s}}{v^{\prime}_{\alpha}\!-\!\ii \Gamma_{\alpha\overline{\alpha}}}
  C_{n,\alpha,s}(x-t)
  \ee^{-\ii (\overline{\epsilon}_{{\rm d}}-\ii \overline{\Gamma}+\frac{\ii }{2}\eta_{s})(x+t)/2}
 \nn\\
&\quad
  -\sum_{P}\sgn(P)e^{(\ell_{P_{1}})}_{\overline{\alpha},k_{P_{1}}}
  \ee^{-\ii k_{P_{1}}t}g^{(\ell_{P_{2}})}_{n,k_{P_{2}}}(x,t).
  \label{eq:2-state_GDQD_e-general_3_2}
\end{eqnarray}
\endnumparts
The initial conditions in Eqs.~\eqref{eq:2-initial-state_GDQD} 
for $e_{n,\alpha}(x,t)$ and $e_{n,\overline{\alpha}}(x,t)$
determine the functions $C_{n,\alpha,s}(x)$ and $C_{n,\alpha,\overline{s}}(x)$ as
\begin{eqnarray}
\fl
 e_{n,\alpha}(x,0+)
 =\sum_{s}C_{n,\alpha,s}(x)
  \ee^{-\ii (\overline{\epsilon}_{{\rm d}}-\ii \overline{\Gamma}+\ii \eta_{s}/2)x/2}
 -\sum_{P}\sgn(P)e^{(\ell_{P_{1}})}_{\alpha,k_{P_{1}}}
  g^{(\ell_{P_{2}})}_{n,k_{P_{2}}}(x,0)=0,
  \nn\\
\fl
 e_{n,\overline{\alpha}}(x,0+)
 =\sum_{s}\frac{\ii \lambda_{\alpha,s}}{v^{\prime}_{\alpha}\!-\!\ii \Gamma_{\alpha\overline{\alpha}}}
  C_{n,\alpha,s}(x)
  \ee^{-\ii (\overline{\epsilon}_{{\rm d}}-\ii \overline{\Gamma}+\ii \eta_{s}/2)x/2}
 \!-\!\sum_{P}\sgn(P)e^{(\ell_{P_{1}})}_{\overline{\alpha},k_{P_{1}}}
  g^{(\ell_{P_{2}})}_{n,k_{P_{2}}}(x,0)=0,
 \nn\\
\fl
\therefore\quad
 C_{n,\alpha,s}(x)
 =\frac{1}{\eta_{s}}
  \sum_{P}\sgn(P)
  g^{(\ell_{P_{1}})}_{n,k_{P_{1}}}(x,0)
  \Big[\lambda_{\alpha,\overline{s}}e^{(\ell_{P_{2}})}_{\alpha,k_{P_{2}}}
  +\ii (v^{\prime}_{\alpha}\!-\!\ii \Gamma_{\alpha\overline{\alpha}})
  e^{(\ell_{P_{2}})}_{\overline{\alpha},k_{P_{2}}}\Big]
 \ee^{\ii (\overline{\epsilon}_{{\rm d}}-\ii \overline{\Gamma}+\ii \eta_{s}/2)x/2}.
 \nn\\
&
 \label{eq:2-state_GDQD_e-C_2}
\end{eqnarray}
It should be noted that, in order to obtain $C_{n,\alpha,s}(x-t)$ for $t>0$ and $x>0$,
we consider both the cases $x-t>0$ and $x-t<0$.
Since  Eq.~\eqref{eq:2-state_GDQD_e-C_2}
determines the function $C_{n,\alpha,s}(x-t)$ only in the case $x-t>0$,
we need another function $D_{n,\alpha,s}(x-t)$ to express 
$C_{n,\alpha,s}(x-t)$ in the case $x-t<0$ as
\begin{eqnarray}
\fl
C_{n,\alpha,s}(x-t)
 =
&D_{n,\alpha,s}(x-t)\theta(t-x)
  \nn\\
\fl
&+\frac{1}{\eta_{s}}
  \sum_{P}\sgn(P)
  g^{(\ell_{P_{1}})}_{n,k_{P_{1}}}(x,t)
  \Big[\lambda_{\alpha,\overline{s}}e^{(\ell_{P_{2}})}_{\alpha,k_{P_{2}}}
  \!+\!\ii (v^{\prime}_{\alpha}\!-\!\ii \Gamma_{\alpha\overline{\alpha}})
  e^{(\ell_{P_{2}})}_{\overline{\alpha},k_{P_{2}}}\Big]
  \nn\\
&\quad\times
  \ee^{\ii (\overline{\epsilon}_{{\rm d}}-\ii \overline{\Gamma}+\ii \eta_{s}/2)(x-t)/2}\theta(x-t).
\end{eqnarray}
By inserting this into Eqs.~\eqref{eq:2-state_GDQD_e-general_3_1} 
and \eqref{eq:2-state_GDQD_e-general_3_2}, 
we obtain
\numparts
\begin{eqnarray}
\fl
e_{m,\alpha}(x,t)
&=\sum_{P}\sgn(P)g^{(\ell_{P_{1}})}_{m,k_{P_{1}}}(x,t)
  e^{(\ell_{P_{2}})}_{\alpha,k_{P_{2}}}(t)
  \nn\\
\fl
&\quad +\sum_{s}D_{m,\alpha,s}(x-t)
 \ee^{-\ii (\overline{\epsilon}_{{\rm d}}-\ii \overline{\Gamma}+\ii \eta_{s}/2)(x+t)/2}
 \theta(t-x),
 \label{eq:2-state_GDQD_e-general_4_1}
 \\
\fl
e_{m,\overline{\alpha}}(x,t)
&=\sum_{P}\sgn(P)g^{(\ell_{P_{1}})}_{m,k_{P_{1}}}(x,t)
   e^{(\ell_{P_{2}})}_{\overline{\alpha},k_{P_{2}}}(t)
  \nn\\
\fl
&\quad
  +\sum_{s}\frac{\ii \lambda_{\alpha,s}}{v^{\prime}_{\alpha}\!-\!\ii \Gamma_{\alpha\overline{\alpha}}}
   D_{m,\alpha,s}(x-t)
   \ee^{-\ii (\overline{\epsilon}_{{\rm d}}-\ii \overline{\Gamma}+\ii \eta_{s}/2)(x+t)/2}
 \theta(t-x).
 \label{eq:2-state_GDQD_e-general_4_2}
\end{eqnarray}
\endnumparts
By using the matching condition for $e_{m,\alpha}(x,t)$ 
in Eq.~\eqref{eq:2-state_GDQD_e-matching}, we have
\begin{eqnarray}
&\sum_{s}D_{m,\alpha,s}(-t)
 \ee^{-\ii (\overline{\epsilon}_{\rm d}-\ii \overline{\Gamma}+\ii \eta_{s}/2)t/2}
 \theta(t)
 =\ii v_{m\overline{\alpha}}X^{(\ell_{1}\ell_{2})}_{\alpha,k_{1}k_{2}}(-t).
\end{eqnarray}
In a similar way, we use the matching condition for $e_{m,\overline{\alpha}}(x,t)$ 
in Eq.~\eqref{eq:2-state_GDQD_e-matching}, obtaining
\begin{eqnarray}
&\sum_{s}\frac{\ii \lambda_{\alpha,s}}{v^{\prime}_{\alpha}-\ii \Gamma_{\alpha\overline{\alpha}}}D_{m,\alpha,s}(-t)
 \ee^{-\ii (\overline{\epsilon}_{\rm d}-\ii \overline{\Gamma}+\ii \eta_{s}/2)t/2}
 =-\ii v_{m\alpha}X^{(\ell_{1}\ell_{2})}_{\alpha,k_{1}k_{2}}(-t).
\end{eqnarray}
By solving the coupled equations for $D_{m,\alpha,+}(-t)$ and $D_{m,\alpha,-}(-t)$, 
we obtain
\begin{eqnarray}
&D_{m,\alpha,s}(-t)
 \ee^{-\ii (\overline{\epsilon}_{\rm d}-\ii \overline{\Gamma}+\ii \eta_{s}/2)t/2}\theta(t)
=-y_{m\alpha,s}X^{(\ell_{1}\ell_{2})}_{\alpha,k_{1}k_{2}}(-t).
\end{eqnarray}
By inserting this into Eqs.~\eqref{eq:2-state_GDQD_e-general_4_1} and
\eqref{eq:2-state_GDQD_e-general_4_2}, we obtain
$e_{m,\alpha}(x,t)$ and $e_{m,\overline{\alpha}}(x,t)$ in the same form
\begin{eqnarray}
e_{m,\alpha}(x,t)
&=\sum_{P}\sgn(P)g^{(\ell_{P_{1}})}_{m,k_{P_{1}}}(x,t)
  e^{(\ell_{P_{2}})}_{\alpha,k_{P_{2}}}(t)
  \nn\\
&\quad
  -\sum_{s}y_{m\alpha,s}X^{(\ell_{1}\ell_{2})}_{\alpha,k_{1}k_{2}}(x-t)
 \ee^{-\ii (\overline{\epsilon}_{\rm d}-\ii \overline{\Gamma}+\ii \eta_{s}/s)x}.
\end{eqnarray}

\noindent
{\bf Step 6)}  Process (iv), (v)$\to$(vi):
The second matching condition in Eqs.~\eqref{eq:2-state_GDQD_g-matching} gives
\begin{eqnarray}
\fl
g_{m_{1}m_{2}}(0+,x,t)
&=\sum_{P}\sgn(P)
 g^{(\ell_{P_{1}})}_{m_{1},k_{P_{1}}}(0+,t)
 g^{(\ell_{P_{2}})}_{m_{2},k_{P_{2}}}(x,t)
 \nn\\
\fl
&\quad
 -\ii \sum_{\alpha, s}v_{m_{1}\alpha}y_{m_{2}\alpha,s}
 X^{(\ell_{1}\ell_{2})}_{\alpha,k_{1}k_{2}}(x-t)
 \ee^{-\ii (\overline{\epsilon}_{\rm d}-\ii \overline{\Gamma}+\ii \eta_{s}/2)x}\theta(x).
\end{eqnarray}
Since $x_{2}-x_{1}>0$ for $0<x_{1}<x_{2}$ in this step, we have
\begin{eqnarray}
\fl
 g_{m_{1}m_{2}}(x_{1},x_{2},t)
&=g_{m_{1}m_{2}}(0+,x_{2}-x_{1},t-x_{1})
  \nn\\
\fl
&=\sum_{P}\sgn(P)
 g^{(\ell_{P_{1}})}_{m_{1},k_{P_{1}}}(x_{1},t)
 g^{(\ell_{P_{2}})}_{m_{2},k_{P_{2}}}(x_{2},t)
 \nn\\
\fl
&\quad 
  -\ii \sum_{\alpha, s}v_{m_{1}\alpha}y_{m_{2}\alpha,s}
  X^{(\ell_{1}\ell_{2})}_{\alpha,k_{1}k_{2}}(x_{2}\!-\! t)
 \ee^{-\ii (\overline{\epsilon}_{\rm d}-\ii \overline{\Gamma}+\ii \eta_{s}/2)x_{21}}\theta(x_{21})\theta(x_{1}).
\end{eqnarray}

\noindent
{\bf Step 7)}  After the anti-symmetrization in the variables $x_{1}$ and $x_{2}$, 
we obtain the wave function $g_{m_{1}m_{2}}(x_{1},x_{2},t)$
in Eqs.~\eqref{eq:2-scattering-state_GDQD_g}.
\begin{flushright}
$\square$
\end{flushright}

We find that the first term of each of the wave functions
in Eqs.~\eqref{eq:2-scattering-state_GDQD_g}, \eqref{eq:2-scattering-state_GDQD_e}
and \eqref{eq:2-scattering-state_GDQD_f}
is a noninteracting part given by the Slater determinant of 
the one-electron wave functions $g^{(\ell)}_{m,k}(x,t)$ and $e^{(\ell)}_{\alpha,k}(t)$
in Eqs.~\eqref{eq:1-scattering-state_GDQD_g} and \eqref{eq:1-scattering-state_GDQD_e}.
The effect of the interdot interaction appears in the second term of each of
Eqs.~\eqref{eq:2-scattering-state_GDQD_g}, \eqref{eq:2-scattering-state_GDQD_e}
and \eqref{eq:2-scattering-state_GDQD_f};
the factor 
$\ee^{-\ii (\overline{\epsilon}_{\rm d}-\ii \overline{\Gamma}+\ii \eta_{s}/2)x_{21}}\theta(x_{21})$
in the second term in Eq.~\eqref{eq:2-scattering-state_GDQD_g}
decays exponentially as the two electrons separate from each other,
which indicates that the two electrons form {\it two-body bound states} after the scattering at the quantum dots.
We also find that the binding strength $\overline{\Gamma}-\mathrm{Re}(\eta_{s}/2)$ of the two-body bound states
corresponds to the resonance pole 
of the one-electron wave functions $g^{(\ell)}_{m,k}(x,t)$ and $e^{(\ell)}_{\alpha,k}(t)$
in Eqs.~\eqref{eq:1-scattering-state_GDQD_g} and \eqref{eq:1-scattering-state_GDQD_e}.

Let us now consider the long-time limit $t\to\infty$ of the time-evolving two-electron scattering  states.
The second term of each of the three wave functions in Eqs.~\eqref{eq:2-scattering-state_GDQD_g},
\eqref{eq:2-scattering-state_GDQD_e} and \eqref{eq:2-scattering-state_GDQD_f}
contains the function
$X^{(\ell_{1}\ell_{2})}_{\alpha,k_{1}k_{2}}(x-t)$ or $X^{(\ell_{1}\ell_{2})}_{\alpha,k_{1}k_{2}}(-t)$,
which includes several terms that decay exponentially in time $t$. 
By using the limit relation
\begin{eqnarray}
\fl
 \lim_{t\to\infty}\ee^{\ii (k_{1}+k_{2})t}X^{(\ell_{1}\ell_{2})}_{\alpha,k_{1}k_{2}}(x-t)
 \nn\\
\fl
 =\sum_{P}\sgn(P)\frac{U^{\prime}}
  {k_{P_{1}}\!+\! k_{P_{2}}\!-\! 2\overline{\epsilon}_{\rm d}\!-\! U^{\prime}\!+\! 2\ii \overline{\Gamma}}
  e^{(\ell_{1})}_{\alpha,k_{1}}e^{(\ell_{2})}_{\overline{\alpha},k_{P_{2}}}
  \ee^{\ii (k_{P_{1}}+k_{P_{2}})x}
 =Z^{(\ell_{1}\ell_{2})}_{\alpha,k_{1}k_{2}}\ee^{\ii (k_{1}+k_{2})x},
 \label{eq:2-scattering-eigenstate_GDQD_Z}
\end{eqnarray}
we find 
\numparts
\begin{eqnarray}
\fl
&\lim_{t\to\infty}\ee^{\ii (k_{1}+k_{2})t}g^{(\ell_{1}\ell_{2})}_{m_{1}m_{2},k_{1}k_{2}}(x_{1},x_{2},t)
 \nn\\
\fl
&=\sum_{P}\sgn(P)
 g^{(\ell_{P_{1}})}_{m_{1},k_{P_{1}}}(x_{1})
 g^{(\ell_{P_{2}})}_{m_{2},k_{P_{2}}}(x_{2})
 \nn\\
\fl
&\quad -\!\ii \sum_{Q\atop \alpha, s}\sgn(Q)v_{m_{Q_{1}}\alpha}y_{m_{Q_{2}}\alpha,s}
  Z^{(\ell_{1}\ell_{2})}_{\alpha,k_{1}k_{2}}
  \ee^{-\ii (\overline{\epsilon}_{\rm d}-\ii \overline{\Gamma}+\ii \eta_{s}/2)x_{Q_{2}Q_{1}}}
  \ee^{\ii Ex_{Q_{2}}}
  \theta(x_{Q_{2}Q_{1}})\theta(x_{Q_{1}})
 \nn\\
\fl
&=: g^{(\ell_{1}\ell_{2})}_{m_{1}m_{2},k_{1}k_{2}}(x_{1},x_{2}),
\label{eq:2-scattering-eigenstate_GDQD_g}
 \\
\fl
&\lim_{t\to\infty}\ee^{\ii (k_{1}+k_{2})t}e^{(\ell_{1}\ell_{2})}_{m\alpha,k_{1}k_{2}}(x,t)
 \nn\\
\fl
&=\sum_{P}\sgn(P)g^{(\ell_{P_{1}})}_{m,k_{P_{1}}}(x)
  e^{(\ell_{P_{2}})}_{\alpha,k_{P_{2}}}
  -\sum_{s}y_{m\alpha,s}Z^{(\ell_{1}\ell_{2})}_{\alpha,k_{1}k_{2}}
  \ee^{-\ii (\overline{\epsilon}_{\rm d}-\ii \overline{\Gamma}+\ii \eta_{s}/2)x}
  \ee^{\ii Ex}
  \theta(x)
 \nn\\
\fl
&=: e^{(\ell_{1}\ell_{2})}_{m\alpha,k_{1}k_{2}}(x),
\label{eq:2-scattering-eigenstate_GDQD_e}
 \\
\fl
&\lim_{t\to\infty}\ee^{\ii (k_{1}+k_{2})t}f^{(\ell_{1}\ell_{2})}_{\alpha\overline{\alpha},k_{1}k_{2}}(t)
 =\sum_{P}\sgn(P)
 e^{(\ell_{P_{1}})}_{\alpha,k_{P_{1}}}
 e^{(\ell_{P_{2}})}_{\overline{\alpha},k_{P_{2}}}
 +Z^{(\ell_{1}\ell_{2})}_{\alpha,k_{1}k_{2}}
 =: f^{(\ell_{1}\ell_{2})}_{\alpha\overline{\alpha},k_{1}k_{2}},
 \label{eq:2-scattering-eigenstate_GDQD_f}
\end{eqnarray}
\endnumparts
which are the stationary two-electron scattering eigenstates
obtained in our previous work~\cite{Nishino-Hatano-Ordonez_16JPC}.
The wave functions of the stationary two-electron scattering eigenstates have 
a resonance pole at $k_{1}+k_{2}=2\overline{\epsilon}_{{\rm d}}+U^{\prime}-2\ii \overline{\Gamma}$ 
on the complex energy plane.

\mathversion{bold}
\section{Time-evolving resonant states}
\mathversion{normal}
\label{sec:time-evolving-resonant-states}

\mathversion{bold}
\subsection{Time-evolving one-body resonant states}
\mathversion{normal}

In Section 3, we presented exact solutions
for the initial conditions in which electrons are absent on the two quantum dots.
We now move onto the solution for the initial condition
in which electrons are present only on the dots.
As a result, we discover new time-evolving states whose wave functions are given in the form
\begin{eqnarray}
 \label{eq:time-evolving-resonant-state}
 \Psi(x,t)=\psi_{R}(x)\ee^{-\ii E_{R}t}\theta(t-x)\quad \mbox{ for } x>0, 
\end{eqnarray}
where $\psi_{R}(x)$ is the wave function of a stationary resonant state
and $E_{R}$ is the resonance energy with an imaginary part.
We refer to the time-evolving states described by Eq.~\eqref{eq:time-evolving-resonant-state}
as {\it time-evolving resonant states}.
Here the step function $\theta(t-x)$ in Eq.~\eqref{eq:time-evolving-resonant-state}, 
which is due to the linear dispersion relation,
restricts the wave function $\Psi(x,t)$ 
to the space interval $0<x<t$;
see Fig.~\ref{fig:EP_PDQD} below.
Hence the time-evolving resonant states are normalizable
even if the corresponding stationary resonant states diverge 
in space~\cite{Hatano-Sasada-Nakamura-Petrosky_08PTP}.
We expect that decaying states are observed in experiments
not as the stationary resonant state but as the time-evolving resonant state.
In the previous work~\cite{Hatano-Sasada-Nakamura-Petrosky_08PTP},
one of the authors showed that the electron number of resonant states is conserved
if the integration interval in the calculation of the electron number
is extended with the electron velocity.
The present time-evolving resonant states clarify 
the meaning of the extension of the integration interval.

First, we consider the time-evolving one-electron state for the initial conditions
\begin{eqnarray}
\label{eq:1-localized-initial-state_GDQD}
&g_{m}(x,0)=0,\quad
 e_{\alpha}(0)=\psi_{\alpha},
\end{eqnarray}
where the constants $\psi_{\alpha}$ for $\alpha=1, 2$ satisfy the normalization condition 
$|\psi_{1}|^{2}+|\psi_{2}|^{2}=1$.

\begin{prop}
{\rm
The set of Schr\"odinger equations~\eqref{eq:Sch-eq_1-state_GDQD_g}
and \eqref{eq:Sch-eq_1-state_GDQD_e}
under the initial conditions~\eqref{eq:1-localized-initial-state_GDQD} 
is solved as follows:
i) For $\eta\neq 0$, the solution is given by
\numparts
\begin{eqnarray}
&g_{m}(x,t)
 =\sum_{\alpha,s}y_{m\alpha,\overline{s}}\,\psi_{\overline{\alpha}}\,
  \ee^{\ii E^{(1)s}_{R}(x-t)}\tilde{\theta}(t-x)\theta(x),
 \label{eq:1-resonant-state_GDQD_g_1}
 \\
&e_{\alpha}(t)
 =\sum_{s}\frac{1}{\eta_{\overline{s}}}
  \bigg[\lambda_{\alpha,\overline{s}}\psi_{\alpha}
  +\ii (v^{\prime}_{\alpha}-\ii \Gamma_{\alpha\overline{\alpha}})\psi_{\overline{\alpha}}\bigg]
  \ee^{-\ii E^{(1)s}_{R}t},
 \label{eq:1-resonant-state_GDQD_e_1}
\end{eqnarray}
\endnumparts
where $E^{(1)\pm}_{R}=\overline{\epsilon}_{\rm d}-\ii \overline{\Gamma}\pm\ii \eta/2$
is the one-body resonance energy.\newline
ii) For $\eta= 0$, the solution is given by
\numparts
\begin{eqnarray}
\fl
&g_{m}(x,t)
 =\ii \sum_{\alpha}v_{m\alpha}((\lambda^{0}_{\alpha}\psi_{\alpha}
  \!+\!\ii (v^{\prime}_{\alpha}\!-\!\ii \Gamma_{\alpha\overline{\alpha}})\psi_{\overline{\alpha}})(t\!-\! x)
  \!-\!\psi_{\alpha})
  \ee^{\ii E^{(1)0}_{R}(x-t)}\tilde{\theta}(t-x)\theta(x),
 \label{eq:1-resonant-state_GDQD_g_1_ep}
 \\
\fl
&e_{\alpha}(t)
 =-((\lambda^{0}_{\alpha}\psi_{\alpha}
  \!+\!\ii (v^{\prime}_{\alpha}\!-\!\ii \Gamma_{\alpha\overline{\alpha}})\psi_{\overline{\alpha}})t\!-\!\psi_{\alpha})
  \ee^{-\ii E^{(1)0}_{R}t},
 \label{eq:1-resonant-state_GDQD_e_1_ep}
\end{eqnarray}
\endnumparts
where $E^{(1)0}_{R}=\overline{\epsilon}_{\rm d}-\ii \overline{\Gamma}$.
}
\end{prop}
\noindent
{\it Proof.}
We employ the systematic construction of three steps proposed 
in Section~\ref{sec:1-time-evolving-scattering-states} for $\eta\neq 0$.
The wave functions for $\eta=0$
in Eqs.~\eqref{eq:1-resonant-state_GDQD_g_1_ep} and \eqref{eq:1-resonant-state_GDQD_e_1_ep}
are obtained by taking the limit $\eta\to 0$ of 
Eqs.~\eqref{eq:1-resonant-state_GDQD_g_1} and \eqref{eq:1-resonant-state_GDQD_g_1}.

\noindent
{\bf Step 1)} In the region of $x<0$ for $t>0$, we have 
$g_{m}(x,t)=g_{m}(x-t,0)=0$ through
the translation invariance in Eq.~\eqref{eq:Sch-eq_1-state_GDQD_g_2}.

\noindent
{\bf Step 2)} Process (i)$\to$(ii): 
Since the inhomogeneous term in Eq.~\eqref{eq:Sch-eq_1-state_GDQD_G}
also vanishes, that is, $G_{\alpha}(t)=0$,
the general solution in Eq.~\eqref{eq:Sch-eq_1-state_GDQD_e_3} is given by
\begin{eqnarray}
&\tilde{e}_{\alpha}(t)
 =\sum_{s}C_{\alpha,s}\ee^{\lambda_{\alpha,s}t},
 \nn\\
\therefore\quad
&e_{\alpha}(t)
 =\ee^{-\ii (\epsilon_{{\rm d}\alpha}-\ii \Gamma_{\alpha\alpha})t}\tilde{e}_{\alpha}(t)
 =\sum_{s}C_{\alpha,s}
  \ee^{\ii (-\epsilon_{{\rm d}\alpha}+\ii \Gamma_{\alpha\alpha}-\ii \lambda_{\alpha,s})t}.
   \label{eq:1-resonant-state_GDQD_e_2}
\end{eqnarray}
By inserting this and $g_{m}(0-,t)=0$ into Eq.~\eqref{eq:Sch-eq_1-state_GDQD_e_1}, we obtain
\begin{eqnarray}
\fl
&\tilde{e}_{\overline{\alpha}}(t)
=\sum_{s}\frac{\ii \lambda_{\alpha,s}}{v^{\prime}_{\alpha}-\ii \Gamma_{\alpha\overline{\alpha}}}
  C_{\alpha,s}
  \ee^{\ii (\Delta\epsilon_{{\rm d}\overline{\alpha}}-\ii \Delta\Gamma_{\overline{\alpha}}-\ii \lambda_{\alpha,s})t},
  \nn\\
\fl
\therefore\quad
&e_{\overline{\alpha}}(t)
 =\ee^{-\ii (\epsilon_{{\rm d}\overline{\alpha}}-\ii \Gamma_{\overline{\alpha}\,\overline{\alpha}})t}\tilde{e}_{\overline{\alpha}}(t)
 =\sum_{s}\frac{\ii \lambda_{\alpha,s}}{v^{\prime}_{\alpha}-\ii \Gamma_{\alpha\overline{\alpha}}}
  C_{\alpha,s}
  \ee^{\ii (-\epsilon_{{\rm d}\alpha}+\ii \Gamma_{\alpha\alpha}-\ii \lambda_{\alpha,s})t}.
  \label{eq:1-resonant-state_GDQD_e_3}
\end{eqnarray}
The initial condition $e_{\alpha}(0)=\psi_{\alpha}$
in Eqs.~\eqref{eq:1-localized-initial-state_GDQD}
determines the constants $C_{\alpha,+}$ and $C_{\alpha,-}$ in Eqs.~\eqref{eq:1-resonant-state_GDQD_e_2}
and \eqref{eq:1-resonant-state_GDQD_e_3} as
\begin{eqnarray}
&\tilde{e}_{\alpha}(0+)
  =C_{\alpha,+}+C_{\alpha,-}=\psi_{\alpha},
  \nn\\
&\tilde{e}_{\overline{\alpha}}(0+)
  =\frac{\ii \lambda_{\alpha,+}}{v^{\prime}_{\alpha}-\ii \Gamma_{\alpha\overline{\alpha}}}\tilde{C}_{\alpha,+}
  +\frac{\ii \lambda_{\alpha,-}}{v^{\prime}_{\alpha}-\ii \Gamma_{\alpha\overline{\alpha}}}\tilde{C}_{\alpha,-}
  =\psi_{\overline{\alpha}},
  \nn\\
\therefore\quad
&C_{\alpha,s}
  =-\frac{1}{\eta_{s}}
   \bigg[\lambda_{\alpha,\overline{s}}\psi_{\alpha}
   +\ii (v^{\prime}_{\alpha}-\ii \Gamma_{\alpha\overline{\alpha}})\psi_{\overline{\alpha}}\bigg].
\end{eqnarray}
By inserting this into Eqs.~\eqref{eq:1-resonant-state_GDQD_e_2} 
and \eqref{eq:1-resonant-state_GDQD_e_3}, we obtain 
\begin{eqnarray}
e_{\alpha}(t)
&=-\sum_{s}\frac{1}{\eta_{s}}
  \bigg[\lambda_{\alpha,\overline{s}}\psi_{\alpha}
  +\ii (v^{\prime}_{\alpha}-\ii \Gamma_{\alpha\overline{\alpha}})\psi_{\overline{\alpha}}\bigg]
  \ee^{-\ii (\overline{\epsilon}_{{\rm d}}-\ii \overline{\Gamma}+\ii \eta_{s}/2)t}.
\end{eqnarray}

\noindent
{\bf Step 3)} Process (i), (ii)$\to$(iii):
Through the matching condition for $g_{m}(x,t)$ in Eq.~\eqref{eq:Sch-eq_1-state_GDQD_g_3}, 
we have
\begin{eqnarray}
g_{m}(0+,t)
&=-\ii \tilde{\theta}(t)\sum_{\alpha}v_{m\alpha}e_{\alpha}(t),
\end{eqnarray}
where, in order to extend the relation to the region $t\leq 0$,
we used the step function $\tilde{\theta}(t)$ which is defined in Eq.~\eqref{eq:step-func}.
Hence, through the translation invariance in Eq.~\eqref{eq:Sch-eq_1-state_GDQD_g_2}, 
we obtain the wave function $g_{m}(x,t)$ for $x>0$ as
\begin{eqnarray}
&g_{m}(x,t)=g_{m}(0+,t-x)
 =-\ii \tilde{\theta}(t-x)\sum_{\alpha, s}v_{m\alpha}e_{\alpha}(t-x).
\end{eqnarray}
\begin{flushright}
$\square$
\end{flushright}

The wave functions $g_{m}(x,t)$ in Eqs.~\eqref{eq:1-resonant-state_GDQD_g_1}
and \eqref{eq:1-resonant-state_GDQD_g_1_ep}
include the step function $\tilde{\theta}(t-x)$, which indicates that
they increase exponentially only inside the space interval $0<x<t$.
This is a result of causality since the electron on the leads travels
with the Fermi velocity $v_{\rm F}=1$ without reflection.
A similar restriction to the time-evolving wave functions due to causality was observed 
in the Friedrichs model~\cite{Petrosky-Ordonez-Prigogine_01PRA}.
It should be noted that the strict step function is due to the precisely linear dispersion
of electrons on the leads; the lower limit of the dispersion would be a branch point and
result in deviations from the purely exponential behavior~\cite{%
Khalfin_68PZETF,Chiu-Sudarshan-Misra_77PRD,Petrosky-Tasaki-Prigogine_91PhysicaA,Petrosky-Ordonez-Prigogine_01PRA,%
Garmon-Petrosky-Simine-Segal_13FortschrPhys,Chakraborty-Sensarma_18PRB,Garmon-Noba-Ordonez-Segal_19PRD}.

Except for the step function $\tilde{\theta}(t-x)$,
the time-evolving state for $\eta\neq 0$
in Eqs.~\eqref{eq:1-resonant-state_GDQD_g_1} and \eqref{eq:1-resonant-state_GDQD_e_1} 
is a superposition of the one-body stationary resonant states 
with the resonance energies $E^{(1)\pm}_{R}$.
Indeed, by using the wave functions $g^{\pm}_{m,R}(x)$ and $e^{\pm}_{\alpha,R}$
of the one-body resonant state, which are explicitly given in \ref{sec:1-ResonantStates},
the wave functions $g_{m}(x,t)$ and $e_{\alpha}(t)$ in 
Eqs.~\eqref{eq:1-resonant-state_GDQD_g_1} and \eqref{eq:1-resonant-state_GDQD_e_1}
are rewritten as
\numparts
\begin{eqnarray}
\label{eq:1-resonant-state_GDQD_g-form_2}
&g_{m}(x,t)
 =\sum_{s}g^{s}_{m,R}(x)\ee^{-\ii E^{(1)s}_{R}t}\tilde{\theta}(t-x),
 \\
\label{eq:1-resonant-state_GDQD_e-form_2}
&e_{\alpha}(t)
 =\sum_{s}e^{s}_{\alpha,R}\ee^{-\ii E^{(1)s}_{R}t},
\end{eqnarray}
\endnumparts
where the constants $\psi_{\alpha}$, which characterize the initial state
in Eqs.~\eqref{eq:1-localized-initial-state_GDQD}, are absorbed into 
the coefficients $g^{s}_{m,R}(x)$ and $e^{s}_{\alpha,R}$.
Since the wave function $g_{m}(x,t)$ is finite in the space interval $0<x<t$,
the time-evolving resonant state is normalizable; see Fig.~\ref{fig:EP_PDQD} below.
Indeed, the time-evolving resonant state in Eqs.~\eqref{eq:1-resonant-state_GDQD_g_1}
and \eqref{eq:1-resonant-state_GDQD_e_1}
is normalized because the initial state 
in Eqs.~\eqref{eq:1-localized-initial-state_GDQD} is normalized
and the particle number is conserved.

The wave functions for $\eta=0$ 
in Eqs.~\eqref{eq:1-resonant-state_GDQD_g_1_ep} and \eqref{eq:1-resonant-state_GDQD_e_1_ep} 
are not in the form of Eqs.~\eqref{eq:1-resonant-state_GDQD_g-form_2}
and \eqref{eq:1-resonant-state_GDQD_e-form_2}
but include an exponential function multiplied by a term linear in $t$.
This is because the two resonance energies $E^{(1)\pm}_{R}$ merge into one at $\eta=0$ and
the corresponding one-body resonant states become parallel to each other, 
which shall be described in \ref{sec:1-ResonantStates}.

\mathversion{bold}
\subsection{Time-evolving two-body resonant states}
\mathversion{normal}

Next, we consider the {\it time-evolving two-body resonant state}
by solving the time-dependent  Schr\"odinger equations~\eqref{eq:Sch-eq_2-state_GDQD_g},
 \eqref{eq:Sch-eq_2-state_GDQD_e} and \eqref{eq:Sch-eq_2-state_GDQD_f}
under the initial conditions 
\begin{eqnarray}
&g_{m_{1}m_{2}}(x_{1},x_{2},0)=0,\quad
 e_{m,\alpha}(x,0)=0,\quad
 f_{\alpha\overline{\alpha}}(0)=(-1)^{\overline{\alpha}},
\label{eq:2-localized-initial-state_GDQD}
\end{eqnarray}
which stand for two electrons localized on the quantum dots.
Recall the anti-symmetry $f_{12}(t)=-f_{21}(t)$
of the wave function of double occupancy.

\begin{prop}
{\rm
For $\eta\neq 0$,
the solution of the set of Schr\"odinger equations~\eqref{eq:Sch-eq_2-state_GDQD_g},
 \eqref{eq:Sch-eq_2-state_GDQD_e} and \eqref{eq:Sch-eq_2-state_GDQD_f}
for the initial conditions~\eqref{eq:2-localized-initial-state_GDQD} 
is given by 
\numparts
\begin{eqnarray}
&g_{m_{1}m_{2}}(x_{1},x_{2},t)
 =\ii \sum_{Q, \alpha, s}(-1)^{\alpha}\sgn(Q)
 v_{m_{Q_{1}}\alpha}y_{m_{Q_{2}}\alpha,s}
 \ee^{\ii E^{(2)}_{R}(x_{Q_{2}}-t)-\ii E^{(1)s}_{R}x_{Q_{2}Q_{1}}}
 \nn\\
&\hspace{93pt}\times
 \tilde{\theta}(t-x_{Q_{2}})\theta(x_{Q_{2}Q_{1}})\theta(x_{Q_{1}}),
\label{eq:2-resonant-state_GDQD_g_1}
  \\
&e_{m,\alpha}(x,t)
 =\sum_{s}(-1)^{\alpha}y_{m\alpha,s}
 \ee^{\ii E^{(2)}_{R}(x-t)-\ii E^{(1)s}_{R}x}\tilde{\theta}(t-x)\theta(x),
 \label{eq:2-resonant-state_GDQD_e_1} 
 \\
\label{eq:2-resonant-state_GDQD_f_1}
&f_{\alpha\overline{\alpha}}(t)
 =(-1)^{\overline{\alpha}}\ee^{-\ii E^{(2)}_{R}t},
\end{eqnarray}
\endnumparts
where $Q=(Q_{1},Q_{2})$ is a permutation of $(1,2)$,
$E^{(1)\pm}_{R}=\overline{\epsilon}_{\rm d}-\ii \overline{\Gamma}+\ii \eta_{s}/2$
is the one-body resonance energy
and $E^{(2)}_{R}=2\overline{\epsilon}_{\rm d}+U^{\prime}-2\ii \overline{\Gamma}$
is the two-body resonance energy.
}
\end{prop}
\noindent
{\it Proof.}
We employ the systematic construction of seven steps proposed 
in Section~\ref{sec:2-time-evolving-scattering-states}.

\noindent
{\bf Step 1)} For $t>0$ and $x_{1}<x_{2}<0$, we find $x_{1}-t<x_{2}-t<0$ in this step.
Then, through the translation invariance in Eq.~\eqref{eq:2-state_GDQD_g-invariance}, 
we find that there is no two-electron incident waves:
\begin{eqnarray}
&g_{m_{1}m_{2}}(x_{1},x_{2},t)
  =g_{m_{1}m_{2}}(x_{1}-t,x_{2}-t,0)
  =0.
\end{eqnarray}

\noindent
{\bf Step 2)} Process (i)$\to$(ii):
Because $g_{m_{1}m_{2}}(x,0-,t)=0$ for $x<0$,
the inhomogeneous term in Eq.~\eqref{eq:2-state_GDQD_e-G} vanishes: 
$G_{m,\alpha}(z)=0$ with $z=(x+t)/2$.
Then we obtain the general solution in Eq.~\eqref{eq:2-state_GDQD_e-general} as
\begin{eqnarray}
&e_{m,\alpha}(x,t)
  =\sum_{s}C_{m,\alpha,s}(x-t)
  \ee^{-\ii (\overline{\epsilon}_{{\rm d}}-\ii \overline{\Gamma}+\ii \eta_{s}/2)(x+t)/2},
  \nn\\
&e_{m,\overline{\alpha}}(x,t)
  =\sum_{s}\frac{\ii \lambda_{\alpha,s}}{v^{\prime}_{\alpha}\!-\!\ii \Gamma_{\alpha\overline{\alpha}}}
  C_{m,\alpha,s}(x-t)
  \ee^{-\ii (\overline{\epsilon}_{{\rm d}}-\ii \overline{\Gamma}+\ii \eta_{s}/2)(x+t)/2},
\end{eqnarray}
where $C_{m,\alpha,s}(x-t)$ for $s=\pm$ are arbitrary functions of the variable $x-t$.
For the initial conditions in Eqs.~\eqref{eq:2-localized-initial-state_GDQD}
for $e_{m,\alpha}(x,t)$ and $e_{m,\overline{\alpha}}(x,t)$,
we obtain $C_{m,\alpha,s}(x)=0$,
that is, $e_{m,\alpha}(x,t)=0$ for $x<0$.

\noindent
{\bf Step 3)} Process (ii)$\to$(iii):
By applying $e_{m,\alpha}(x,t)=0$ for $x<0$
to the solution in Eq.~\eqref{eq:2-state_GDQD_f-solution}
for $f_{\alpha\overline{\alpha}}(t)$, we have
\begin{eqnarray}
f_{\alpha\overline{\alpha}}(t)
&=C_{\alpha}\ee^{-\ii (2\overline{\epsilon}_{\rm d}+U^{\prime}-2\ii \overline{\Gamma})t}.
\end{eqnarray}
Under the initial condition $f_{\alpha\overline{\alpha}}(0)=(-1)^{\overline{\alpha}}$ 
in Eqs.~\eqref{eq:2-localized-initial-state_GDQD}, we have
\begin{eqnarray}
&f_{\alpha\overline{\alpha}}(0+)=C_{\alpha}=(-1)^{\overline{\alpha}},
  \nn\\
\therefore\quad
&f_{\alpha\overline{\alpha}}(t)
  =(-1)^{\overline{\alpha}}\ee^{-\ii (2\overline{\epsilon}_{\rm d}+U^{\prime}-2\ii \overline{\Gamma})t}.
\end{eqnarray}

\noindent
{\bf Step 4)} Process (i), (ii)$\to$(iv):
For $x_{1}<0<x_{2}$, we have $x_{1}-x_{2}<0$ in this step.
Then the first matching condition in Eqs.~\eqref{eq:2-state_GDQD_g-matching} gives
\begin{eqnarray}
g_{m_{1}m_{2}}(x_{1},x_{2},t)
&=g_{m_{1}m_{2}}(x_{1}-x_{2},0+,t-x_{2})
=0.
\end{eqnarray}

\noindent
{\bf Step 5)} Process (iii), (iv)$\to$(v):
Because $g_{m_{1}m_{2}}(0-,x,t)=0$ for $x>0$,
the inhomogeneous term in Eq.~\eqref{eq:2-state_GDQD_e-G}
also vanishes: $G_{m,\alpha}(z)=0$ with $z=(x+t)/2$.
Then, by using the solution in Eq.~\eqref{eq:2-state_GDQD_e-general}, we have
\begin{eqnarray}
\label{eq:2-state_GDQD_e-RS}
&e_{m,\alpha}(x,t)
  =\sum_{s}\tilde{C}_{m,\alpha,s}(x-t)
 \ee^{-\ii (\overline{\epsilon}_{{\rm d}}-\ii \overline{\Gamma}+\ii \eta_{s}/2)(x+t)/2},
  \nn\\
&e_{m,\overline{\alpha}}(x,t)
  =\sum_{s}\frac{\ii \lambda_{\alpha,s}}{v^{\prime}_{\alpha}\!-\!\ii \Gamma_{\alpha\overline{\alpha}}}
  \tilde{C}_{m,\alpha,s}(x-t)
  \ee^{-\ii (\overline{\epsilon}_{{\rm d}}-\ii \overline{\Gamma}+\ii \eta_{s}/2)(x+t)/2},
\end{eqnarray}
where $\tilde{C}_{m,\alpha,s}(x-t)$ for $s=\pm$ are arbitrary functions of the variable $x-t$.
From the initial conditions~\eqref{eq:2-localized-initial-state_GDQD}, we have
\begin{eqnarray}
&e_{m,\alpha}(x,0)
 =\sum_{s}\tilde{C}_{m,\alpha,s}(x)
 \ee^{-\ii (\overline{\epsilon}_{{\rm d}}-\ii \overline{\Gamma}+\ii \eta_{s}/2)x/2}
 =0,
 \nn\\
&e_{m,\overline{\alpha}}(x,0)
 =\sum_{s}\frac{\ii \lambda_{\alpha,s}}{v^{\prime}_{\alpha}\!-\!\ii \Gamma_{\alpha\overline{\alpha}}}
  \tilde{C}_{m,\alpha,s}(x)
  \ee^{-\ii (\overline{\epsilon}_{{\rm d}}-\ii \overline{\Gamma}+\ii \eta_{s}/2)x/2}
  =0,
 \nn\\
\therefore\quad
&\tilde{C}_{m,\alpha,s}(x)=0\quad \mbox{ for } x>0,
\end{eqnarray}
which determines $\tilde{C}_{m,\alpha,s}(x-t)=0$ for $x-t>0$.
In order to express the function $\tilde{C}_{m,\alpha,s}(x-t)$ for $x-t<0$,
we introduce another function $D_{m,\alpha,s}(x-t)$ and put
\begin{eqnarray}
&\tilde{C}_{m,\alpha,s}(x-t)=D_{m,\alpha,s}(x-t)\theta(t-x).
\end{eqnarray}
By inserting this into the solution in Eqs.~\eqref{eq:2-state_GDQD_e-RS}, we have
\begin{eqnarray}
e_{m,\alpha}(x,t)
&=\sum_{s}D_{m,\alpha,s}(x-t)
 \ee^{-\ii (\overline{\epsilon}_{{\rm d}}-\ii \overline{\Gamma}+\ii \eta_{s}/2)(x+t)/2}
 \theta(t-x),
 \\
e_{m,\overline{\alpha}}(x,t)
&=\sum_{s}\frac{\ii \lambda_{\alpha,s}}{v^{\prime}_{\alpha}\!-\!\ii \Gamma_{\alpha\overline{\alpha}}}
 D_{m,\alpha,s}(x-t)
 \ee^{-\ii (\overline{\epsilon}_{{\rm d}}-\ii \overline{\Gamma}+\ii \eta_{s}/2)(x+t)/2}
 \theta(t-x).
\end{eqnarray}
The matching condition in Eq.~\eqref{eq:2-state_GDQD_e-matching}
for $e_{m,\alpha}(x,t)$ gives
\begin{eqnarray}
\fl
&\sum_{s}D_{m,\alpha,s}(-t)
 \ee^{-\ii (\overline{\epsilon}_{{\rm d}}-\ii \overline{\Gamma}+\ii \eta_{s}/2)t/2}
 =(-1)^{\overline{\alpha}}\ii v_{m\overline{\alpha}}
  \ee^{-\ii (2\overline{\epsilon}_{\rm d}+U^{\prime}-2\ii \overline{\Gamma})t},
  \nn\\
\fl
&\sum_{s}\frac{\ii \lambda_{\alpha,s}}{v^{\prime}_{\alpha}-\ii \Gamma_{\alpha\overline{\alpha}}}D_{m,\alpha,s}(-t)
 \ee^{-\ii (\overline{\epsilon}_{{\rm d}}-\ii \overline{\Gamma}+\ii \eta_{s}/2)t/2}
 =(-1)^{\alpha}\ii v_{m\alpha}
  \ee^{-\ii (2\overline{\epsilon}_{\rm d}+U^{\prime}-2\ii \overline{\Gamma})t}.
\end{eqnarray}
By solving the coupled equations for $D_{m,\alpha,s}(-t)$, we obtain
\begin{eqnarray}
&D_{m,\alpha,s}(-t)
 \ee^{-\ii (\overline{\epsilon}_{\rm d}-\ii \overline{\Gamma}+\ii \eta_{s}/2)t/2}
=(-1)^{\alpha}y_{m\alpha,s}\psi_{\alpha\overline{\alpha}}
 \ee^{-\ii (2\overline{\epsilon}_{\rm d}+U^{\prime}-2\ii \overline{\Gamma})t}.
\end{eqnarray}
By replacing $t(>0)$ with $t-x(>0)$ in the variable of $D_{m,\alpha,s}(-t)$, 
we have 
\begin{eqnarray}
&D_{m,\alpha,s}(x-t)
 \ee^{-\ii (\overline{\epsilon}_{\rm d}-\ii \overline{\Gamma}+\ii \eta_{s}/2)(x+t)/2}\tilde{\theta}(t-x)
 \nn\\
&=(-1)^{\alpha}\tilde{\theta}(t-x)y_{m\alpha,s}
 \ee^{\ii (2\overline{\epsilon}_{\rm d}+U^{\prime}-2\ii \overline{\Gamma})(x-t)
 -\ii (\overline{\epsilon}_{\rm d}-\ii \overline{\Gamma}+\ii \eta_{s}/2)x},
\end{eqnarray}
which gives
\begin{eqnarray}
\fl
e_{m,\alpha}(x,t)
=(-1)^{\alpha}\tilde{\theta}(t-x)\sum_{s}y_{m\alpha,s}
 \ee^{\ii (\overline{\epsilon}_{\rm d}+U^{\prime}-\ii \overline{\Gamma}-\ii \eta_{s}/2)x
 -\ii (2\overline{\epsilon}_{\rm d}+U^{\prime}-2\ii \overline{\Gamma})t}\theta(x).
\end{eqnarray}

\noindent
{\bf Step 6)} Process (iv), (v)$\to$(vi):
By using the second matching condition in Eqs.~\eqref{eq:2-state_GDQD_g-matching},
we have
\begin{eqnarray}
\fl
g_{m_{1}m_{2}}(0+,x,t)
=\ii \tilde{\theta}(t-x)\sum_{\alpha, s}(-1)^{\alpha}v_{m_{1}\alpha}y_{m_{2}\alpha,s}
 \ee^{\ii (\overline{\epsilon}_{\rm d}+U^{\prime}-\ii \overline{\Gamma}-\ii \eta_{s}/2)x
 -\ii (2\overline{\epsilon}_{\rm d}+U^{\prime}-2\ii \overline{\Gamma})t}\theta(x).
\end{eqnarray}
Because $x_{2}-x_{1}>0$ for $0<x_{1}<x_{2}$, we have
\begin{eqnarray}
\fl
 g_{m_{1}m_{2}}(x_{1},x_{2},t)
 =g_{m_{1}m_{2}}(0+,x_{2}-x_{1},t-x_{1})
  \nn\\
\fl
 =\ii \sum_{\alpha, s}(-1)^{\alpha}v_{m_{1}\alpha}y_{m_{2}\alpha,s}
 \ee^{\ii (\overline{\epsilon}_{\rm d}+U^{\prime}-\ii \overline{\Gamma}-\ii \eta_{s}/2)x_{21}
 +\ii (2\overline{\epsilon}_{\rm d}+U^{\prime}-2\ii \overline{\Gamma})(x_{1}-t)}
 \tilde{\theta}(t-x_{2})\theta(x_{21})\theta(x_{1}).
\end{eqnarray}

\noindent
{\bf Step 7)} After the anti-symmetrization with respect to the variables $x_{1}$ and $x_{2}$,
we obtain the wave function $g_{m_{1}m_{2}}(x_{1},x_{2},t)$ in the case $x_{2}<x_{1}$.
\begin{flushright}
$\square$
\end{flushright}

The time-evolving two-body resonant state
is normalizable since the wave functions $g_{m_{1}m_{2}}(x_{1},x_{2},t)$ is finite
in the space region $0<x_{1}, x_{2}<t$,
and $e_{m,\alpha}(x,t)$ is also finite in the space interval $0<x<t$, 
which is due to the step functions that imply causality.
We also find that the two electrons in the two-body resonant state
form two-body bound states with the binding strength $\overline{\Gamma}-\mathrm{Re}(\eta_{s}/2)$,
which is the imaginary part of the resonance energy $E^{(1)s}_{R}$. 

By using the wave functions $g_{m_{1}m_{2},R}(x_{1},x_{2})$, $e_{m,\alpha,R}(x)$ 
and $f_{\alpha\overline{\alpha},R}$ of the two-body stationary resonant state given 
in \ref{sec:2-ResonantStates},
the wave functions for $\eta\neq 0$ in Eqs.~\eqref{eq:2-resonant-state_GDQD_g_1},
\eqref{eq:2-resonant-state_GDQD_e_1} and \eqref{eq:2-resonant-state_GDQD_f_1} are rewritten as
\begin{eqnarray}
\label{eq:2-resonant-state_GDQD_g_2}
&g_{m_{1}m_{2}}(x_{1},x_{2},t)
 =g_{m_{1}m_{2},R}(x_{1},x_{2})
 \ee^{-\ii E^{(2)}_{R}t}\tilde{\theta}(t-x_{1})\tilde{\theta}(t-x_{2}),
 \\
\label{eq:2-resonant-state_GDQD_e_2}
&e_{m,\alpha}(x,t)
 =e_{m,\alpha,R}(x)\ee^{-\ii E^{(2)}_{R}t}\tilde{\theta}(t-x),
 \\
\label{eq:2-resonant-state_GDQD_f_2}
&f_{\alpha\overline{\alpha}}(t)
 =f_{\alpha\overline{\alpha},R}\ee^{-\ii E^{(2)}_{R}t}.
\end{eqnarray}
Hence we refer to this time-evolving two-electron state as {\it a time-evolving two-body resonant state}.
The relation $E^{(2)}_{R}=U^{\prime}+E^{(1)+}_{R}+E^{(1)-}_{R}$ among the resonance energies 
shows that, at $U^{\prime}=0$,  the time-evolving two-body resonant state
is reduced to a combination of the two time-evolving one-body resonant states
with different resonance energies $E^{(1)+}_{R}$ and $E^{(1)-}_{R}$.
We remark that, in a way similar to the one-electron case, the wave functions for $\eta=0$,
which are obtained by taking the limit $\eta\to 0$ of Eqs.~\eqref{eq:2-resonant-state_GDQD_g_1},
\eqref{eq:2-resonant-state_GDQD_e_1} and \eqref{eq:2-resonant-state_GDQD_f_1},
include an exponential function multiplied by a term linear in $t$.

\subsection{Existence and survival probabilities }
\label{sec:probabilities}

The explicit wave functions of the time-evolving states enable us to
calculate the time-dependence of the existence probability of electrons on the leads
and the survival probability of electrons on the quantum dots.
First, we investigate the existence probability 
under the initial condition of electrons localized on the quantum dots.
By using the wave function $g_{m}(x,t)$ in Eq.~\eqref{eq:1-resonant-state_GDQD_g_1}
of the time-evolving one-body resonant state, we obtain
the existence probability distribution of an electron on the lead $m$ at time $t$ as
\begin{eqnarray}
\label{eq:existence-prob_1}
\fl
P^{(1)}_{m}(x,t)
=|g_{m}(x,t)|^{2}
=\Big|\sum_{\alpha,s}y_{m\alpha,s}\psi_{\overline{\alpha}}\,
   \ee^{(\overline{\Gamma}+\eta_{s}/2)(x-t)}\Big|^{2}
   \tilde{\theta}(t-x)\theta(x).
\end{eqnarray}
Similarly, by using the wave function $e_{m,\alpha}(x,t)$ 
in Eq.~\eqref{eq:2-resonant-state_GDQD_e_1} 
of the time-evolving two-body resonant state,
we obtain the existence probability distribution of an electron on the lead $m$
in the presence of another electron on the quantum dots at time $t$ as
\begin{eqnarray}
\label{eq:existence-prob_2}
\fl
 P^{(2)}_{m{\rm d}}(x,t)
 =\sum_{\alpha}|e_{m,\alpha}(x,t)|^{2}
=\sum_{\alpha}\Big|\sum_{s}y_{m\alpha,s}
 \ee^{((\overline{\Gamma}+\eta_{s}/2)x-2\overline{\Gamma}t)}\Big|^{2}\tilde{\theta}(t-x)\theta(x).
\end{eqnarray}
We note that the existence probability $P^{(2)}_{m{\rm d}}(x,t)$ in Eq.~\eqref{eq:existence-prob_2}
does not depend on the interaction $U^{\prime}$
since the interaction $U^{\prime}$ appears only in the real part of the two-body resonance energy 
$E^{(2)}_{R}=2\overline{\epsilon}_{\rm d}+U^{\prime}-2\ii \overline{\Gamma}$ 
and hence in the phase factor of the wave function $e_{m,\alpha}(x,t)$ 
in Eq.~\eqref{eq:2-resonant-state_GDQD_e_2}.

We remind that the existence probability $P^{(1)}_{m}(x,t)$ in Eq.~\eqref{eq:existence-prob_1} describes  
an electron on the lead $m$ in the one-electron case,
while $P^{(2)}_{m{\rm d}}(x,t)$ in Eq.~\eqref{eq:existence-prob_2} describes an electron on the lead $m$ 
and another electron on the quantum dots in the two-electron case.
We find from Eqs.~\eqref{eq:existence-prob_1} and \eqref{eq:existence-prob_2} 
that, at fixed time $t$, both the existence probabilities $P^{(1)}_{m}(x,t)$ 
and $P^{(2)}_{m{\rm d}}(x,t)$ increase exponentially 
with the same strength $2\overline{\Gamma}+|\mathrm{Re}(\eta)|$
in the space interval $0<x<t$~\cite{Petrosky-Ordonez-Prigogine_01PRA}.
Thus, even in the two-electron case, the exponential increase in space is dominated
by the one-body resonance energy $E^{(1)s}_{R}$, which has been seen in the two-body bound state
of the two-electron scattering wave function in Eq.~\eqref{eq:2-scattering-state_GDQD_g}.
On the other hand, at a fixed position $x$,
the existence probability $P^{(1)}_{m}(x,t)$ decays exponentially in time 
with the inverse relaxation time $2\overline{\Gamma}-|\mathrm{Re}(\eta)|$,
while $P^{(2)}_{m{\rm d}}(x,t)$ decays exponentially in time with $4\overline{\Gamma}$.
The exponential decay in time in the two-electron case is dominated 
by the two-body resonance energy $E^{(2)}_{R}$.

\begin{figure}[t]
\begin{center}
{
\begin{picture}(340,225)(0,0)
 \put(-10,-10){\includegraphics[width=330pt]{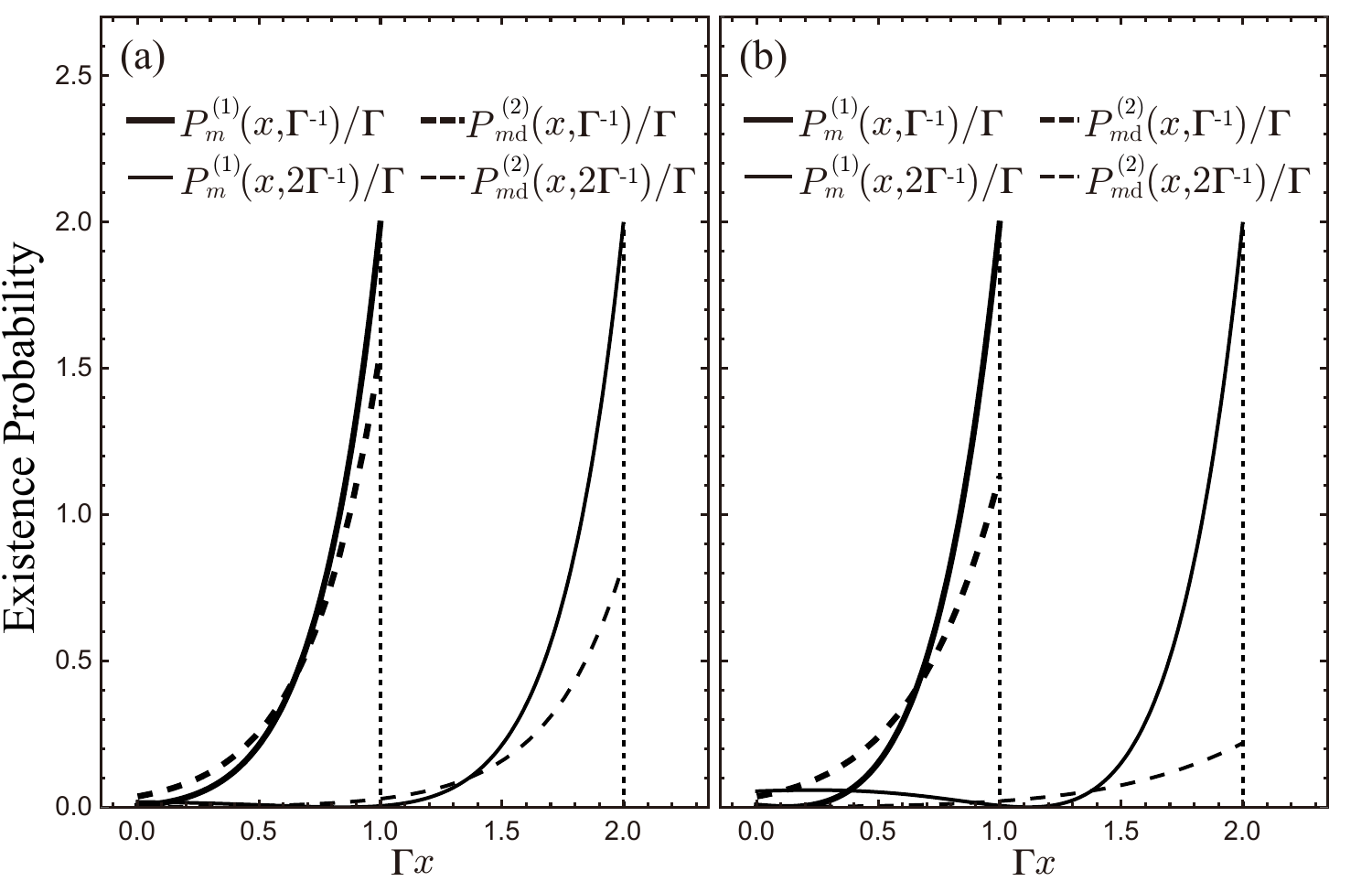}}
\end{picture}
}
\end{center}
\caption{
Existence probability distributions of the time-evolving one- and two-body resonant states
in the case of the parallel-coupled DQD system
with $v_{\ell\alpha}=v\in\mathbb{R}$, $v^{\prime}_{\alpha}=0$ 
for (a) $\epsilon_{{\rm d}1}=-\epsilon_{{\rm d}2}=0.8\Gamma$
and (b) $\epsilon_{{\rm d}1}=-\epsilon_{{\rm d}2}=1.2\Gamma$.
The solid lines indicate the existence probability
$P^{(1)}_{m}(x,t)/\Gamma$ of an electron on the lead $m$
in the one-electron case,
and the dashed lines indicate the existence probability
$P^{(2)}_{m{\rm d}}(x,t)/\Gamma$ of an electron on the lead $m$
and another electron on the quantum dots in the two-electron case.
}
\label{fig:EP_PDQD}

\end{figure}

For numerical demonstration,
we consider the parallel-coupled double quantum-dot system
by setting $v_{m\alpha}=v(\in\mathbb{R})$, 
$v^{\prime}_{\alpha}=0$ and $\epsilon_{{\rm d}1}=-\epsilon_{{\rm d}2}=\epsilon$,
in which we have $\eta=2(\Gamma^{2}-\epsilon^{2})^{1/2}$ with $\Gamma=v^{2}$.
Here we find an exceptional point $\eta=0$ at $\Gamma=|\epsilon|$.
For $\Gamma>|\epsilon|$, the one-body resonance energies 
$E^{(1)\pm}_{R}=\epsilon-\ii \Gamma\pm\ii \eta/2$ have different imaginary parts,
while for $\Gamma<|\epsilon|$, they have different real parts.
We take the initial state in Eq.~\eqref{eq:1-localized-initial-state_GDQD} 
with $\psi_{1}=\psi_{2}=1/\sqrt{2}$
for the existence probability $P^{(1)}_{m}(x,t)$,
while the initial state for $P^{(2)}_{m{\rm d}}(x,t)$ is Eq.~\eqref{eq:2-localized-initial-state_GDQD}.
Then the one- and two-electron existence probabilities 
$P^{(1)}_{m}(x,t)$ and $P^{(2)}_{m{\rm d}}(x,t)$ 
in Eqs.~\eqref{eq:existence-prob_1} and \eqref{eq:existence-prob_2} 
inside the space interval $0<x<t$ are explicitly calculated as
\begin{eqnarray}
\label{eq:1-existence-prob_PDQD}
\fl 
&P^{(1)}_{m}(x,t)
\!=\!
 \cases{
  2\Big|\frac{2\Gamma}{\eta}\sinh\Big(\frac{\eta}{2}(t-x)\Big)
  -\cosh\Big(\frac{\eta}{2}(t-x)\Big)\Big|^{2}
  \Gamma\ee^{-2\Gamma(t-x)}\hspace{-15pt}
  & for $\Gamma>|\epsilon|$, \\
  2\big|\Gamma(t-x)-1\big|^{2}\Gamma\ee^{-2\Gamma(t-x)}\hspace{-15pt}
  & for $\Gamma=|\epsilon|$, \\
  2\Big|\frac{2\Gamma}{\zeta}\sin\Big(\frac{\zeta}{2}(t-x)\Big)
  -\cos\Big(\frac{\zeta}{2}(t-x)\Big)\Big|^{2}
  \Gamma\ee^{-2\Gamma(t-x)}\hspace{-15pt}
  & for $\Gamma<|\epsilon|$, \\
 }
 \\
\label{eq:2-existence-prob_PDQD}
\fl 
&P^{(2)}_{m{\rm d}}(x,t)
\!=\!
 \cases{\!
  \sum_{\alpha}\Big|\frac{2(\ii \epsilon_{{\rm d}\alpha}-\Gamma)}{\eta}\sinh\Big(\frac{\eta}{2}x\Big)
  -\cosh\Big(\frac{\eta}{2}x\Big)\Big|^{2}
  \Gamma\ee^{-2\Gamma(2t-x)}\hspace{-15pt}
  & for $\Gamma>|\epsilon|$, \\
  \sum_{\alpha}\big|\big(\ii \epsilon_{{\rm d}\alpha}-\Gamma\big)x-1\big|^{2}\Gamma\ee^{-2\Gamma(2t-x)}
  & for $\Gamma=|\epsilon|$, \\
  \sum_{\alpha}\Big|\frac{2(\ii \epsilon_{{\rm d}\alpha}-\Gamma)}{\zeta}\sin\Big(\frac{\zeta}{2}x\Big)
  -\cos\Big(\frac{\zeta}{2}x\Big)\Big|^{2}
  \Gamma\ee^{-2\Gamma(2t-x)}\hspace{-15pt}
  & for $\Gamma<|\epsilon|$. \\
 }
\end{eqnarray}
Here we put $\zeta=2(\epsilon^{2}-\Gamma^{2})^{1/2}$ for $\Gamma<|\epsilon|$.
The existence probabilities for $\Gamma=|\epsilon|$ are obtained 
by taking the limit $\eta\to 0$ of Eqs.~\eqref{eq:existence-prob_1} and \eqref{eq:existence-prob_2}.
Figure \ref{fig:EP_PDQD} shows the existence probability distributions 
$P^{(1)}_{m}(x,t)$ and $P^{(2)}_{m{\rm d}}(x,t)$ 
for (a) $\epsilon_{{\rm d}1}=-\epsilon_{{\rm d}2}=0.8\Gamma$ and
(b) $\epsilon_{{\rm d}1}=-\epsilon_{{\rm d}2}=1.2\Gamma$;
the solid lines indicate the rescaled existence probability 
$P^{(1)}_{m}(x,t)/\Gamma$ and the dashed lines indicate $P^{(2)}_{m{\rm d}}(x,t)/\Gamma$.
The thick lines correspond to the existence probability at time $t=\Gamma^{-1}$
and the thin lines correspond to that at time $t=2\Gamma^{-1}$.
In the case (b), the initial decrease of $P^{(2)}_{m{\rm d}}(x,2\Gamma)$ is
a part of oscillation of the existence probabilities before the exponential increase,
which is due to the interference of the two resonance energies $E^{(1)\pm}_{R}$
with different real parts and is expressed by the trigonometric functions 
in Eqs.~\eqref{eq:1-existence-prob_PDQD} and \eqref{eq:2-existence-prob_PDQD}.
As Eqs.~\eqref{eq:1-existence-prob_PDQD} and \eqref{eq:2-existence-prob_PDQD} indicate,
for fixed $x$, the two-electron existence probability $P^{(2)}_{m{\rm d}}(x,t)$
with the relaxation time $4\Gamma$ decays more rapidly than $P^{(1)}_{m}(x,t)$
with $2\Gamma-|\mathrm{Re}(\eta)|$ as a function of $t$.

Next we investigate the survival probabilities of electrons on the quantum dots.
By using the wave function $e_{\alpha}(t)$ in Eq.~\eqref{eq:1-resonant-state_GDQD_e_1},
we obtain the one-electron survival probability $Q^{(1)}(t)$ on the quantum dots at time $t$ as
\begin{eqnarray}
\label{eq:1-survival-prob}
\fl
Q^{(1)}(t)
 =\sum_{\alpha}|e_{\alpha}(t)|^{2}
 =\sum_{\alpha}\Big|\sum_{s}\frac{1}{\eta_{\overline{s}}}
  \big(\lambda_{\alpha,\overline{s}}\psi_{\alpha}
  +\ii (v^{\prime}_{\alpha}-\ii \Gamma_{\alpha\overline{\alpha}})\psi_{\overline{\alpha}}\big)
  \ee^{-(\overline{\Gamma}-\eta_{s}/2)t}\Big|^{2},
\end{eqnarray}
which decays exponentially with the inverse lifetime $2\overline{\Gamma}-|\mathrm{Re}(\eta)|$.
On the other hand, the two-electron survival probability $Q^{(2)}(t)$ on the quantum dots is obtained
from the wave function of double occupancy $f_{12}(t)$ in Eq.~\eqref{eq:2-resonant-state_GDQD_f_1} 
at time $t$ as
\begin{eqnarray}
\label{eq:2-survival-prob}
 Q^{(2)}(t)=|f_{12}(t)|^{2}=\ee^{-4\overline{\Gamma}t},
\end{eqnarray}
which decays exponentially with the inverse lifetime $4\overline{\Gamma}$.
Thus  the lifetime of the survival probability $Q^{(1)}(t)$ on the quantum dots
is equal to the relaxation time of the existence probability $P^{(1)}_{m}(x,t)$ on the leads,
while the lifetime of $Q^{(2)}(t)$ is equal to the relaxation time of $P^{(2)}_{m{\rm d}}(x,t)$.

\begin{figure}[t]
\begin{center}
{
\begin{picture}(310,200)(0,0)
 \put(0,-10){\includegraphics[width=300pt]{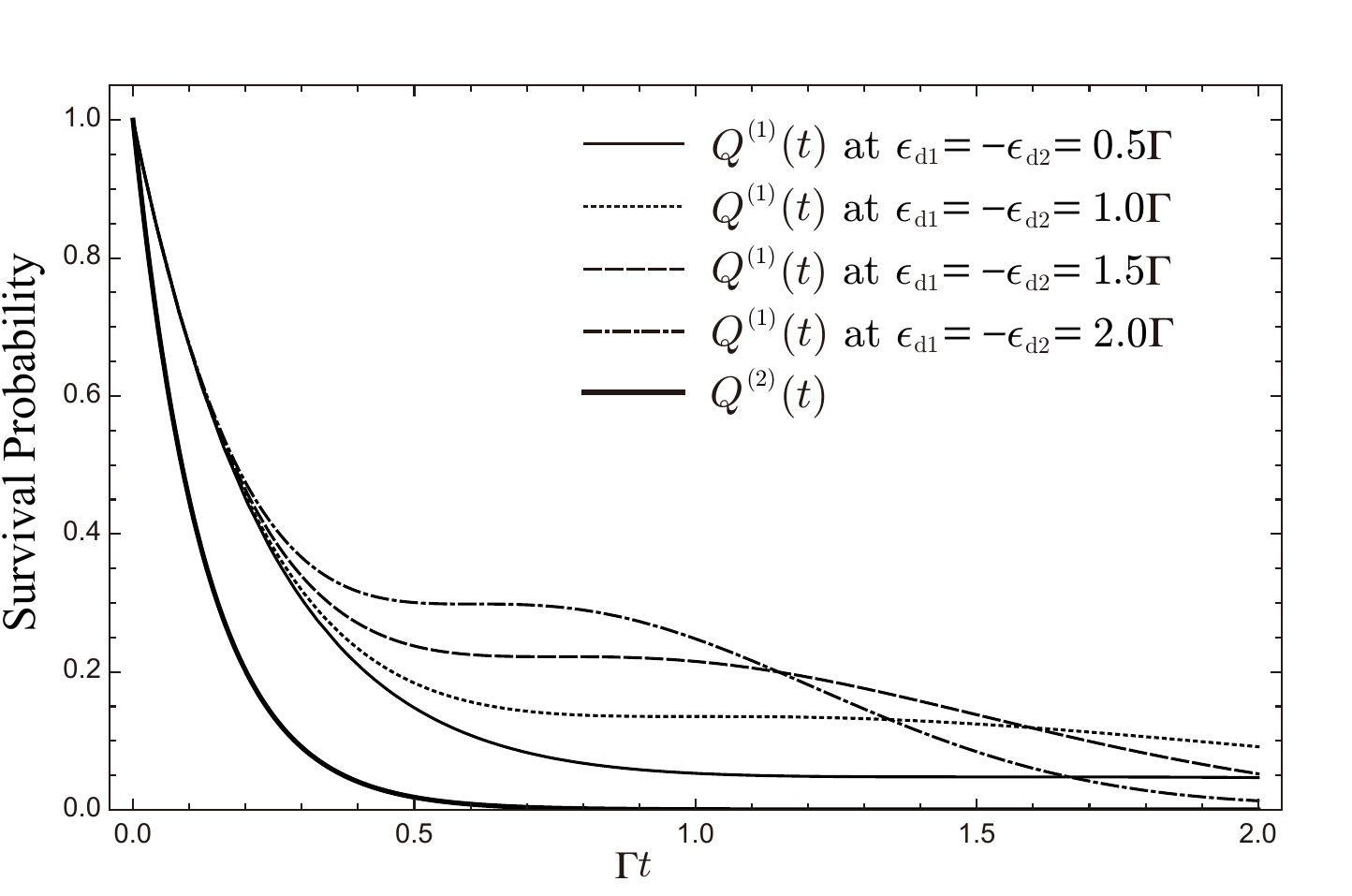}}
\end{picture}
}
\end{center}
\caption{
Survival probabilities of the time-evolving one- and two-body resonant states
in the case of the parallel-coupled double quantum-dot system
with $v_{\ell\alpha}=v\in\mathbb{R}$, $v^{\prime}_{\alpha}=0$ and $\epsilon_{{\rm d}1}=-\epsilon_{{\rm d}2}$.
The solid thin line indicates the survival probabilities 
$Q^{(1)}(t)$ of the time-evolving one-body resonant states 
at $\epsilon_{{\rm d}1}=0.5\Gamma$,
the dotted line indicates that  at $\epsilon_{{\rm d}1}=1.0\Gamma$,
the dashed line indicates that  at $\epsilon_{{\rm d}1}=1.5\Gamma$,
the chain line indicates that at $\epsilon_{{\rm d}1}=2.0\Gamma$
and the solid thick line indicates the survival probability
$Q^{(2)}(t)$ of the time-evolving two-body resonant states.
}
\label{fig:SP_PDQD}
\end{figure}

Let us analyze the case of the parallel-coupled double quantum-dot system
with $v_{m\alpha}=v(\in\mathbb{R})$, $v^{\prime}_{\alpha}=0$ 
and $\epsilon_{{\rm d}1}=-\epsilon_{{\rm d}2}=\epsilon$ again.
Recall that $\eta=2(\Gamma^{2}-\epsilon^{2})^{1/2}$ for $\Gamma>|\epsilon|$
and $\zeta=2(\epsilon^{2}-\Gamma^{2})^{1/2}$ for $\Gamma<|\epsilon|$ with $\Gamma=v^{2}$.
We take the initial state in Eq.~\eqref{eq:1-localized-initial-state_GDQD}
with $\psi_{1}=\psi_{2}=1/\sqrt{2}$
for the one-electron survival probability $Q^{(1)}(t)$,
which is the same as that of the existence probabilities
in Eq.~\eqref{eq:1-existence-prob_PDQD}.
Then we have
\begin{equation}
\label{eq:1-survival-prob_PDQD}
\fl
 Q^{(1)}(t)
\!=\!
 \cases{\!
  \frac{1}{2}\sum_{\alpha}\Big|
  \frac{2(\ii \epsilon_{{\rm d}\alpha}+\Gamma)}{\eta}
  \sinh\Big(\frac{\eta}{2}t\Big)-\cosh\Big(\frac{\eta}{2}t\Big)
  \Big|^{2}
  \ee^{-2\Gamma t}\hspace{-15pt}
  & for $\Gamma>|\epsilon|$, \\
  \!
  \frac{1}{2}\sum_{\alpha}
  \big|\big(\ii \epsilon_{{\rm d}\alpha}+\Gamma\big)t-1\big|^{2}\ee^{-2\Gamma t}
  & for $\Gamma=|\epsilon|$, \\
  \!
  \frac{1}{2}\sum_{\alpha}\Big|
  \frac{2(\ii \epsilon_{{\rm d}\alpha}+\Gamma)}{\zeta}
  \sin\Big(\frac{\zeta}{2}t\Big)-\cos\Big(\frac{\zeta}{2}t\Big)
  \Big|^{2}
  \ee^{-2\Gamma t}\hspace{-15pt}
  & for $\Gamma<|\epsilon|$. \\
 }
\end{equation}
Figure~\ref{fig:SP_PDQD} shows the one-electron survival probabilities $Q^{(1)}(t)$  
for $\epsilon=0.5\Gamma, \Gamma, 1.5\Gamma, 2\Gamma$
and the two-electron survival probability $Q^{(2)}(t)$ in Eq.~\eqref{eq:2-survival-prob},
which does not depend on the energy level $\epsilon$.
Due to the difference of their lifetime,
the two-electron survival probability $Q^{(2)}(t)$ decays 
more rapidly than $Q^{(1)}(t)$ as a function of $t$.
We also observe that, for $\Gamma<|\epsilon|$, 
the survival probability $Q^{(1)}(t)$ oscillates in time,
which is expressed by the trigonometric functions in Eqs.~\eqref{eq:1-survival-prob_PDQD}.
The oscillation in time is a result of the interference of 
the two resonance energies $E^{(1)\pm}_{R}$with different real parts.

\section{Concluding remarks}
\label{sec:concluding-remarks}

We have constructed exact time-evolving states of the open double quantum-dot system.
Their purely exponential decay and strict step-function behavior
are consequences of the unbounded linear dispersion relation of electrons on the leads.
For noninteracting cases, it was shown~\cite{Gurvitz_15PhysScr} that
the time-evolving scattering states are characterized 
by a non-Markovian integro-differential equation,
because of which the behavior of the survival probability of unstable states 
deviates from the purely exponential decay
usually in an extremely short time and in an extremely long time
if the dispersion is bounded below~\cite{Khalfin_68PZETF,%
Chiu-Sudarshan-Misra_77PRD,Petrosky-Tasaki-Prigogine_91PhysicaA,Petrosky-Ordonez-Prigogine_01PRA,%
Garmon-Petrosky-Simine-Segal_13FortschrPhys,Chakraborty-Sensarma_18PRB,Garmon-Noba-Ordonez-Segal_19PRD}.
In the case of the linear dispersion relation, the integro-differential equation
is reduced to a Markovian differential equation and hence produces exponential functions
without deviations.
We have also shown that the exponential decay of the time-evolving states
leads to that of the existence probability of an electron on the leads
and the survival probability of localized electrons on the quantum dots.

The systematic construction of time-evolving states that we have proposed
can be readily extended to the cases of three or more electrons.
We have found that the two-body bound states in the two-electron scattering states
correspond to the one-body resonance poles of the one-electron scattering states.
In a similar way, three-body bound states appear in three-electron scattering states
due to the two-body resonance poles of the two-electron scattering states.
Each term of the many-electron scattering states is characterized by
the arrangement of the two-body and the three-body bound states,
which is an essential difference from the interacting resonant-level models
with only two-body bound states~\cite{%
Nishino-Imamura-Hatano_09PRL,Nishino-Imamura-Hatano_11PRB,Nishino-Hatano-Ordonez_15PRB,%
Culver-Andrei_21PRB_1,Culver-Andrei_21PRB_2,Culver-Andrei_21PRB_3}.
The time-evolving many-electron scattering states shall enable us
to analyze time-dependent electric current across the quantum dots
for the system under bias voltages.

\section*{Acknowledgments}

The authors thank Prof.~T.~Y.~Petrosky, Prof.~G.~Ordonez and Prof.~S.~Garmon for helpful comments.
The present study is partially supported by Japan Society for the Promotion of Science (JSPS) Grants Numbers 
19H00658, 21H01005, 22H01140.

\appendix
\section{Resonant states and Siegert conditions}
\label{sec:ResonantStates}

\subsection{One-body resonant states}
\label{sec:1-ResonantStates}

In this appendix, we study the stationary resonant states which are eigenstates of 
the time-independent Schr\"odinger equation with complex energy eigenvalues.
First, we consider the one-electron case.
By assuming the time-dependent parts of the one-electron wave functions
of the set of time-dependent  Schr\"odinger equations~\eqref{eq:Sch-eq_1-state_GDQD_g}
and \eqref{eq:Sch-eq_1-state_GDQD_e} as
\begin{eqnarray}
\label{eq:t-indep-Sch-eq_1-state_GDQD_0}
 g_{m}(x,t)=\ee^{-\ii Et}g_{m}(x),\quad
 e_{\alpha}(t)=\ee^{-\ii Et}e_{\alpha}
\end{eqnarray}
with an energy eigenvalue $E$
and inserting them into Eqs.~\eqref{eq:Sch-eq_1-state_GDQD_g_3}
and \eqref{eq:Sch-eq_1-state_GDQD_e_1},
we have
\begin{eqnarray}
\label{eq:t-indep-Sch-eq_1-state_GDQD_g_1}
&g_{m}(0+)-g_{m}(0-)=-\ii \sum_{\alpha}v_{m\alpha}e_{\alpha},
 \\
\label{eq:t-indep-Sch-eq_1-state_GDQD_e_1}
&(E-\epsilon_{{\rm d}\alpha}+\ii \Gamma_{\alpha\alpha})e_{\alpha}
 -\big(v^{\prime}_{\alpha}-\ii \Gamma_{\alpha\overline{\alpha}}\big)e_{\overline{\alpha}}
 =\sum_{m}v_{m\alpha}^{\ast}g_{m}(0-).
\end{eqnarray}

To define the resonant states,
we employ the Siegert boundary conditions that impose no incident wave
on the wave functions~\cite{Siegert_39PR} as
\begin{eqnarray}
 g_{m}(x)=0 \mbox{ for } x<0.
\end{eqnarray}
Hereafter we put the suffix $R$ on the wave functions as $g_{m,R}(x)$ and $e_{\alpha,R}$
to denote the wave functions on which the Siegert boundary conditions are imposed.
Then Eq.~\eqref{eq:t-indep-Sch-eq_1-state_GDQD_e_1} with $g_{m, R}(0-)=0$
is reduced to the following eigenvalue problem of a non-Hermite matrix: 
\begin{equation}
\label{eq:non-Hermite_1-state_GDQD_1}
\left(
\begin{array}{cc}
 \epsilon_{{\rm d}\alpha}-\ii \Gamma_{\alpha\alpha} 
 & v^{\prime}_{\alpha}-\ii \Gamma_{\alpha\overline{\alpha}} \\
 v^{\prime}_{\overline{\alpha}}-\ii \Gamma_{\overline{\alpha}\alpha} 
 & \epsilon_{{\rm d}\overline{\alpha}}-\ii \Gamma_{\overline{\alpha}\,\overline{\alpha}}
\end{array}
\right)
\left(
\begin{array}{c}
 e_{\alpha,R} \\
 e_{\overline{\alpha},R}
\end{array}
\right)
=E
\left(
\begin{array}{c}
 e_{\alpha,R} \\
 e_{\overline{\alpha},R}
\end{array}
\right).
\end{equation}
The non-Hermite matrix on the left-hand side
is referred to as an effective Hamiltonian of the open system~\cite{%
Hatano-Sasada-Nakamura-Petrosky_08PTP,Hatano-Ordonez_19Book}.
It is characteristic to the systems with linear dispersion relations
that the imaginary term $-\ii \Gamma_{\alpha\beta}$,
so-called the self-energy of the leads,  in each matrix element
is independent of energy $E$. 
By diagonalizing the non-Hermite matrix, we obtain resonance energies 
\begin{eqnarray}
 E^{(1)\pm}_{R}=\overline{\epsilon}_{\rm d}-\ii \overline{\Gamma}+\frac{\ii }{2}\eta_{\pm},
\end{eqnarray}
where $\eta_{\pm}$ has been defined in Eqs.~\eqref{eq:lambda-eta}.
The corresponding eigenvector $(e^{s}_{\alpha,R}, e^{s}_{\overline{\alpha},R})$
for $s=\pm$ must satisfy the relation
\begin{eqnarray}
&\frac{e^{s}_{\alpha,R}}{e^{s}_{\overline{\alpha},R}}
 =\frac{v^{\prime}_{\alpha}-\ii \Gamma_{\alpha\overline{\alpha}}}{\ii \lambda_{\alpha,s}}
 =\frac{\ii \lambda_{\overline{\alpha},s}}{v^{\prime}_{\overline{\alpha}}-\ii \Gamma_{\overline{\alpha}\alpha}}.
\end{eqnarray}
It is readily found that, at the exceptional points satisfying $\eta=0$,
the two eigenvalues $E^{(1)s}_{R}$ for $s=\pm$ 
merge into the one eigenvalue $\overline{\epsilon}_{\rm d}-\ii \overline{\Gamma}$ 
while the eigenvectors $(e^{+}_{\alpha,R}, e^{+}_{\overline{\alpha},R})$ and
$(e^{-}_{\alpha,R}, e^{-}_{\overline{\alpha},R})$
become parallel to each other.

In order to relate the eigenvectors to the time-evolving resonant state for $\eta\neq 0$
in Eqs.~\eqref{eq:1-resonant-state_GDQD_g_1} and \eqref{eq:1-resonant-state_GDQD_e_1}, 
we take the expression
\begin{eqnarray}
 \label{eq:1-t-indep-resonant-state_GDQD_e}
e^{s}_{\alpha,R}
&=\frac{1}{\eta_{\overline{s}}}\big(\psi_{\alpha}\lambda_{\alpha,\overline{s}}
  +\psi_{\overline{\alpha}}\,\ii (v^{\prime}_{\alpha}-\ii \Gamma_{\alpha\overline{\alpha}})\big)
\end{eqnarray}
with the constants $\psi_{\alpha}$ and $\psi_{\overline{\alpha}}$.
We note that the ambiguity of the above expression of the eigenvectors is only a constant multiple
due to the relation
$\lambda_{\alpha,s}\lambda_{\overline{\alpha},s}
  =-(v^{\prime}_{\alpha}-\ii \Gamma_{\alpha\overline{\alpha}})
     (v^{\prime}_{\overline{\alpha}}-\ii \Gamma_{\overline{\alpha}\alpha})$.
By using the matching condition in Eq.~\eqref{eq:t-indep-Sch-eq_1-state_GDQD_g_1}, 
we obtain the wave function $g_{m,R}(x)$ of the resonant state as
\begin{eqnarray}
&g^{s}_{m,R}(0+)
 =\sum_{\alpha}\psi_{\overline{\alpha}}\frac{1}{\eta_{\overline{s}}}
 \big(\ii \lambda_{\alpha,s}v_{m\overline{\alpha}}
  +(v^{\prime}_{\alpha}-\ii \Gamma_{\alpha\overline{\alpha}})v_{m\alpha}\big)
  =\sum_{\alpha}\psi_{\overline{\alpha}}y_{m\alpha,\overline{s}},
  \nn\\
\therefore\quad
&g^{s}_{m,R}(x)
 =\sum_{\alpha}\psi_{\overline{\alpha}}y_{m\alpha,\overline{s}}\ee^{\ii E^{(1)s}_{R}x}\theta(x).
 \label{eq:1-t-indep-resonant-state_GDQD_g}
\end{eqnarray}
By inserting the wave function $g^{s}_{m,R}(x)$ in Eq.~\eqref{eq:1-t-indep-resonant-state_GDQD_g} 
and $e^{s}_{\alpha,R}$ in Eq.~\eqref{eq:1-t-indep-resonant-state_GDQD_e}
into Eqs.~\eqref{eq:t-indep-Sch-eq_1-state_GDQD_0},
we obtain the one-body stationary resonant state.
The imaginary part $\overline{\Gamma}-\mathrm{Re}(\eta_{s}/2)$ 
of the resonance energy $E^{(1)s}_{R}$
corresponds to the inverse lifetime of the resonant state.

It should be noted that the resonance energy $E^{(1)s}_{R}$ has appeared as a resonance pole
of the one-electron scattering eigenfunctions in Eqs.~\eqref{eq:1-scattering-eigenstate_GDQD_g}
and \eqref{eq:1-scattering-eigenstate_GDQD_e}.
We remark that the above wave functions $g^{s}_{m,R}(x)$ and $e^{s}_{\alpha,R}$ of the one-body resonant state
are also obtained by taking a residue 
at the resonance pole $k=\overline{\epsilon}_{\rm d}-\ii \overline{\Gamma}+\ii \eta_{s}/2$
of the one-electron stationary scattering eigenfunctions $g^{(\ell)}_{m,k}(x)$ and $e^{(\ell)}_{\alpha,k}$
in Eqs.~\eqref{eq:1-scattering-eigenstate_GDQD_g} and \eqref{eq:1-scattering-eigenstate_GDQD_e}.

\subsection{Two-body resonant states}
\label{sec:2-ResonantStates}

Next, we consider the two-body stationary resonant states.
In a way similar to the one-body case, 
we assume the time-dependent parts of the two-electron wave functions 
of the time-dependent  Schr\"odinger equations~\eqref{eq:Sch-eq_2-state_GDQD_g},
\eqref{eq:Sch-eq_2-state_GDQD_e} and \eqref{eq:Sch-eq_2-state_GDQD_f} as
\begin{eqnarray}
\label{eq:t-indep-Sch-eq_2-state_GDQD_0}
\fl
&g_{m_{1}m_{2}}(x_{1},x_{2},t)=\ee^{-\ii Et}g_{m_{1}m_{2}}(x_{1},x_{2}),\quad
 \nn\\
&e_{m,\alpha}(x,t)=\ee^{-\ii Et}e_{m,\alpha}(x),\quad
 f_{\alpha\overline{\alpha}}(t)=\ee^{-\ii Et}f_{\alpha\overline{\alpha}}.
\end{eqnarray}
By inserting these into Eq.~\eqref{eq:2-state_GDQD_f-PDE_1},
we have
\begin{eqnarray}
\label{eq:t-indep-Sch-eq_2-state_GDQD_f}
&(E-2\overline{\epsilon}_{\rm d}-U^{\prime}+2\ii \overline{\Gamma})
 f_{\alpha\overline{\alpha}}
 =\sum_{m, \beta}
  (-)^{\alpha+\beta}v^{\ast}_{m\beta}e_{m,\overline{\beta}}(0-).
\end{eqnarray}
We recall that the wave function $e_{m,\alpha}(x)$ describes two electrons,
one of which is on the quantum dots and another is on the leads;
see Eq.~\eqref{eq:2-state_GDQD}.

As an extension of the Siegert boundary conditions~\cite{Siegert_39PR} to the two-electron case,
we impose the conditions that there is neither a one-electron nor a two-electron incident wave: 
\begin{eqnarray}
&g_{m_{1}m_{2}}(x_{1},x_{2})=0\quad \mbox{ for } x_{1}<0 \mbox{ or } x_{2}<0,
 \nn\\
&e_{m,\alpha}(x)=0\quad \mbox{ for } x<0.
\end{eqnarray}
Then, by applying $e_{m,\alpha}(0-)=0$ to Eq.~\eqref{eq:t-indep-Sch-eq_2-state_GDQD_f},
we obtain the {\it two-body} resonance energy 
\begin{eqnarray}
 E^{(2)}_{R}=2\overline{\epsilon}_{\rm d}+U^{\prime}-2\ii \overline{\Gamma},
\end{eqnarray}
which depends on the strength $U^{\prime}$ of the interdot interaction.

Since there is no restriction on the wave function $f_{\alpha\overline{\alpha},R}$,
we take $f_{\alpha\overline{\alpha},R}=(-1)^{\overline{\alpha}}$.
Through the systematic construction of scattering eigenstates proposed 
in the previous paper~\cite{Nishino-Hatano-Ordonez_16JPC},
we have
\begin{eqnarray}
\fl
 g_{m_{1}m_{2},R}(x_{1},x_{2})
 =\ii \sum_{Q,\alpha,s}(-)^{\alpha}\sgn(Q)
  v_{m_{Q_{1}}\alpha}y_{m_{Q_{2}}\alpha,s}
  \ee^{\ii (U^{\prime}+E^{(1)\overline{s}}_{R})x_{Q_{2}Q_{1}}+\ii E^{(2)}_{R}x_{Q_{1}}}
  \theta(x_{Q_{2}Q_{1}})\theta(x_{Q_{1}}),
 \nn\\
\fl
 e_{m,\alpha,R}(x)
 =(-)^{\alpha}\sum_{s}y_{m\alpha,s}
  \ee^{\ii (U^{\prime}+E^{(1)\overline{s}}_{R})x}\theta(x).
\end{eqnarray}
By inserting these into Eqs.~\eqref{eq:t-indep-Sch-eq_2-state_GDQD_0},
we obtain the two-body stationary resonant state.
The imaginary part $2\overline{\Gamma}$ of the resonance energy $E^{(2)}_{R}$
gives the inverse lifetime of the two-body resonant state.

We note that the two-body resonance energy has appeared as the resonance pole
of the two-electron scattering eigenfunctions in Eqs.~\eqref{eq:2-scattering-eigenstate_GDQD_g},
\eqref{eq:2-scattering-eigenstate_GDQD_e} and \eqref{eq:2-scattering-eigenstate_GDQD_f}
with Eq.~\eqref{eq:2-scattering-eigenstate_GDQD_Z}.
We remark that the above wave functions $g_{m_{1}m_{2},R}(x_{1},x_{2})$, $e_{m,\alpha,R}(x)$
and $f_{\alpha\overline{\alpha},R}$ 
of the two-body resonant state are also obtained by taking a residue 
at the resonance pole $E=2\overline{\epsilon}_{\rm d}+U^{\prime}-2\ii \overline{\Gamma}$
of the stationary two-electron scattering eigenfunctions 
$g^{(\ell_{1}\ell_{2})}_{m_{1}m_{2},k_{1}k_{2}}(x_{1},x_{2})$,
$e^{(\ell_{1}\ell_{2})}_{m\alpha,k_{1}k_{2}}(x)$ 
and $f^{(\ell_{1}\ell_{2})}_{\alpha\overline{\alpha},k_{1}k_{2}}$
in Eqs.~\eqref{eq:2-scattering-eigenstate_GDQD_g},
\eqref{eq:2-scattering-eigenstate_GDQD_e} and \eqref{eq:2-scattering-eigenstate_GDQD_f}.

\section*{References}

\begin{thebibliography}{10}
\bibitem{Cronenwett-Oosterkamp-Kouwenhoven_98Science}
Cronenwett S M, Oosterkamp T H and Kouwenhoven L P 
1998 {\it Science} {\bf 281} 540
\bibitem{GoldhaberGordon-Shtrikman-Mahalu-Abusch-Magder-Meirav-Kastner_98Nature}
Goldhaber-Gordon D, Shtrikman H, Mahalu D, Abusch-Magder D, Meirav U and Kastner M A 
1998 {\it Nature (London)} {\bf 391} 156
\bibitem{VanDerWiel-DeFranceschi-Fujisawa-Elzerman-Tarucha-Kouwenhoven_00Science}
van der Wiel W G, De Franceschi S, Fujisawa T, Elzerman J M, Tarucha S and Kouwenhoven L P 
van der Wiel W G \etal
2000 {\it Science} {\bf 289} 2105
\bibitem{Landauer_57IBMJRD}
Landauer R
1957 {\it IBM J. Res. Dev.} {\bf 1} 223
\bibitem{Buttiker_86PRL}
B\"uttiker M 
1986 {\it Phys. Rev. Lett.} {\bf 57} 1761
\bibitem{Bagwell-Orlando_98PRB}
Bagwell P F and Orlando T P 
1989 {\it Phys. Rev.} B {\bf 40} 1456
\bibitem{Meir-Wingreen_92PRL}
Meir Y and Wingreen N S 
1992 {\it Phys. Rev. Lett.} {\bf 68} 2512
\bibitem{Meir-Wingreen-Lee_93PRL}
Meir Y and Wingreen N S 
1993 {\it Phys. Rev. Lett.} {\bf 70} 2601
\bibitem{Wingreen-Meir_94PRB}
Wingreen N S and Meir Y 1994 
{\it Phys. Rev.} B {\bf 49} 11040
\bibitem{Costi-Hewson-Zlatic_94JPCM}
T.~A.~Costi, A.~C.~Hewson and V.~Zlati\'c, J.~Phys.~Condens.~Matter {\bf 6}, 2519 (1994).
\bibitem{Nishino-Imamura-Hatano_09PRL}
Nishino A, Imamura T and Hatano N 
2009 {\it Phys. Rev. Lett.} {\bf 102} 146803
\bibitem{Nishino-Imamura-Hatano_11PRB}
Nishino A, Imamura T and Hatano N 
2011 {\it Phys. Rev.} B {\bf 83} 035306
\bibitem{Nishino-Hatano-Ordonez_15PRB}
Nishino A, Hatano N and Ordonez G 
2015 {\it Phys. Rev.} B {\bf 91} 045140
\bibitem{Doyon_07PRL}
Doyon B 
2007 {\it Phys. Rev. Lett.} {\bf 99} 076806
\bibitem{Boulat-Saleur-Schmitteckert_08PRL}
Boulat E, Saleur H and Schmitteckert P 
2008 {\it Phys. Rev. Lett.} {\bf 101} 140601
\bibitem{Golub_07PRB}
Golub A
2007 {\it Phys. Rev.} B {\bf 76} 193307
\bibitem{Karrasch-Andergassen-Pletyukhov-Schuricht-Borda-Meden-Schoeller_10EL}
Karrasch C, Andergassen S, Pletyukhov M, Schuricht D, Borda L, Meden V and Schoeller H, 
2010 {\it Europhys. Lett.} {\bf 90} 30003
\bibitem{Imamura-Nishino-Hatano_09PRB}
Imamura T, Nishino A and Hatano N 
2009 {\it Phys. Rev.} B {\bf 80} 245323
\bibitem{Nishino-Imamura-Hatano_12JPC}
Nishino A, Imamura T and Hatano N 
2012 \JPCS {\bf 343} 012087
\bibitem{Nishino-Hatano-Ordonez_16JPC}
Nishino A, Hatano N and Ordonez G
2016 \JPCS {\bf 670} 012038
\bibitem{Gamow_28ZPhysA}
Gamow V G
1928 {\it Z. Phys.} A {\bf 51} 204
\bibitem{Siegert_39PR}
Siegert A J F 
1939 {\it Phys. Rev.} {\bf 56} 750
\bibitem{Peierls_59PRSLA}
Peierls R E 
1959 {\it Proc. Roy. Soc. London} A {\bf 253} 16
\bibitem{Hokkyo_65PTP}
Hokkyo N 
1965 {\it Prog. Theor. Phys.} {\bf 33} 1116
\bibitem{Berggren_68NPA}
Berggren T
1968 {\it Nucl. Phys.} A {\bf 109} 265
\bibitem{Romo_68NPA}
Romo W 
1968 {\it Nucl. Phys.} A {\bf 116} 618
\bibitem{Berggren_70PLB}
Berggren T 
1970 {\it Phys. Lett.} B {\bf 33} 547
\bibitem{Lind_93PRA}
Lind P 
1993 {\it Phys. Rev. C} {\bf 47} 1903
\bibitem{Hatano-Sasada-Nakamura-Petrosky_08PTP}
Hatano N, Sasada K, Nakamura H and Petrosky T
2008 {\it Prog. Theor. Phys.} {\bf 119} 187
\bibitem{Hatano-Ordonez_19Book}
Hatano N and Ordonez G 
2018 
{\it Fano Resonances in Optics and Microwaves} (Cham: Springer) P.~357
\bibitem{Hatano-Ordonez_19Entropy}
Hatano N and Ordonez G 
2019 {\it Entropy} {\bf 21} No. 380
\bibitem{Tanaka-Kawakami_05PRB}
Tanaka Y and Kawakami N
2005 {\it Phys. Rev.} B {\bf 72} 085304
\bibitem{Culver-Andrei_21PRB_1}
Culver A B and Andrei N 
2021 {\it Phys. Rev.} B {\bf 103} L201103
\bibitem{Culver-Andrei_21PRB_2}
Culver A B and Andrei N 
2021 {\it Phys. Rev.} B {\bf 103} 195106
\bibitem{Culver-Andrei_21PRB_3}
Culver A B and Andrei N 
2021 {\it Phys. Rev.} B {\bf 103} 195107
\bibitem{Khalfin_68PZETF}
Khalfin L A 
1968 {\it Pis'ma Zh. Eksp. Teor. Fiz.} {\bf 8} 106 [{\it JETP Letters} {\bf 8} 65]
\bibitem{Chiu-Sudarshan-Misra_77PRD}
Chiu C B, Sudarshan E C G and Misra B
1977 {\it Phys. Rev.} D {\bf 16} 520
\bibitem{Petrosky-Tasaki-Prigogine_91PhysicaA}
Petrosky T, Tasaki S and Prigogine I
1991 {\it Physica} A {\bf 170} 306
\bibitem{Petrosky-Ordonez-Prigogine_01PRA}
Petrosky T, Ordonez G and Prigogine I
2001 {\it Phys. Rev.} A {\bf 64} 062101
\bibitem{Garmon-Petrosky-Simine-Segal_13FortschrPhys}
Garmon S, Petrosky T, Simine L and Segal D
2013 {\it Fortschr. Phys.} {\bf 61} 261
\bibitem{Chakraborty-Sensarma_18PRB}
Chakraborty A and Sensarma R
2018 {\it Phys. Rev.} B {\bf 97} 104306
\bibitem{Garmon-Noba-Ordonez-Segal_19PRD}
Garmon S, Noba K, Ordonez G and Segal D
2019 {\it Phys. Rev.} A {\bf 99} 010102
\bibitem{Chowdhury-Chattopadhyay_23AQT}
Chowdhury B N and Chattopadhyay S
2023 {\it Adv. Quantum Technol.} {\bf 6} 2200072
\bibitem{Gurvitz_15PhysScr}
Gurvitz S
2015 {\it Phys. Scr.} {\bf 2015} 014013
\bibitem{Hewson}
Hewson A C 1993
{\it The Kondo Problem to Heavy Fermions}
(Cambridge University Press) P.~138
\end{thebibliography}

\end{document}